\documentclass[conference]{IEEEtran}
\usepackage{url}
\usepackage{subfigure}
\usepackage{framed}
\usepackage{pdflscape}
\usepackage{longtable}
\usepackage{doi}
\usepackage{amsmath,amssymb,amsfonts}
\usepackage{algorithmic}
\usepackage{graphicx}
\usepackage{textcomp}
\usepackage{xcolor}
\usepackage{scalefnt}
\usepackage[numbers, square, comma, sort&compress]{natbib}
\usepackage{subfigure}
\usepackage{url}
\usepackage{subfigure}
\usepackage{framed}
\usepackage{changepage}
\def\BibTeX{{\rm B\kern-.05em{\sc i\kern-.025em b}\kern-.08em
    T\kern-.1667em\lower.7ex\hbox{E}\kern-.125emX}}

\begin{document}

\title{Cloud Network Slicing: A Systematic Mapping Study  from Scientific Publications\\
\thanks{Identify applicable funding agency here. If none, delete this.}
}

\author{\IEEEauthorblockN{Leandro C. de Almeida}
\IEEEauthorblockA{\textit{Academic Unit of Informatics} \\
\textit{Federal Institute of Para\'iba}\\
Jo\~ao Pessoa-PB, Brazil \\
leandro.almeida@ifpb.edu.br}
\and
\IEEEauthorblockN{Paulo Ditarso Maciel Jr.}
\IEEEauthorblockA{\textit{Academic Unit of Informatics} \\
\textit{Federal Institute of Para\'iba}\\
Jo\~ao Pessoa-PB, Brazil \\
paulo.maciel@ifpb.edu.br}
\and
\IEEEauthorblockN{F\'abio L. Verdi}
\IEEEauthorblockA{\textit{Dept. of Computer Science} \\
\textit{Federal University of S\~ao Carlos}\\
Sorocaba-SP, Brazil \\
verdi@ufscar.br}
}

\maketitle

\begin{abstract}
Cloud Network Slicing is a new research area that brings together cloud computing and network slicing in an end-to-end environment. In this context, understanding the existing scientific contributions and gaps is crucial to driving new research in this field. This article presents a complete quantitative analysis of scientific publications on the Cloud Network Slicing, based on a systematic mapping study. The results indicate the situation of the last ten years in the research area, presenting data such as industry involvement, most cited articles, most active researchers, publications over the years, main places of publication, as well as well-developed areas and gaps. Future guidelines for scientific research are also discussed.
\end{abstract}


In the past, \citet{Peterson:2003} brought a disruptive view to computer network architectures: for the first time the term \textbf{Slice} was used in the context of computer networks. Since then, the topic of slicing has evolved so that, like any other hot topic, there has been a surge of scientific publications in recent years. Several Standards Developing Organizations (SDO) have also been creating documents to define what slice is in the context of telecom operators, cloud and network providers. 

In this context, we mention some examples of well-known SDOs working on the definition of slicing: ETSI (European Telecommunications Standards Institute) \cite{ETSI:2018}, IETF (Internet Engineering Task Force) \cite{IETF:2019}, 3GPP (3rd Generation Partnership Project) \cite{3GPP:2018}, NGMN (Next Generation Mobile Networks)\cite{NGMN:2016} and ITU-T (ITU Telecommunication Standardization Sector) \cite{ITU-T:2017}.

Although there is no unified definition on the concept of slice, several articles have been published in recent years in this area \cite{Richart:2019, Foukas:2017, Rost:2017, Zhou:2016}. In the field of computer networks, slice takes advantage of technologies like SDN (Software Defined Networking) and NFV (Network Functions Virtualization), allowing to build a programmable and dynamic structure on demand. Coupled with the concept of cloud computing, Slice enables the creation of a more complex architecture (\textbf{CNS - Cloud Network Slicing}) that encompasses network and cloud technologies, enabling new services \cite{Soenen:2017}. 

In this sense, CNS can be defined as the process that enables isolated end-to-end and on-demand networking abstractions, which: (a) contain both cloud and network resources, and (b) are independently controlled, managed and orchestrated \cite{Maciel:2019}. 

Critical communications, V2X (Vehicular-to-X), Massive IoT (Internet of Things) and eMBB (enhanced Mobile Broadband) are examples of new technologies that can benefit from cloud network slicing. Different services have different requirements, such as very high throughput, large connection density or ultra-low latency. In this sense, CNS must be able to support services with different characteristics, according to the defined SLA (Service Level Agreement) \cite{Kaloxylos:2018}.

Some papers in the literature present open challenges in the CNS context \cite{Soenen:2017, Zhang:2017, Maciel:2019, Xie:2019}. Topics such as monitoring, elasticity, isolation, security, QoS (Quality of Service), open interfaces (standardization), resource discovery, and mobility haven't been completely addressed yet.

In a relatively new research area, like CNS, craft a research agenda is difficult for researchers. This is probably due to the need for an accurate investigation into a research problem \cite{Miles:2017}. In this sense, \emph{evidence-based research} could assist researchers to identify well-developed areas and/or critical gaps. 

An example of evidence-based research is a systematic mapping study, which is a type of secondary study focused in discovering research gaps and trends \cite{Petersen:2008}. Unlike systematic literature reviews, which focus on synthesizing scientific evidence, systematic maps are primarily concerned with structuring a research area \cite{PETERSEN20151}.

Given the current importance of the CNS concept, this work presents results about an in-depth systematic mapping study. The main contribution of this study is a holistic view, represented by a bubble plot, about scientific contributions in cloud network slicing. Furthermore, other results are presented such as publications over the years, industry involvement, main researchers, main conferences/journals and most cited papers. From the results, we present evidences of the challenges that are still open and future directions for the CNS area.

The remainder of this article is arranged as follows. In Section \ref{sec:background}, we present the background and the fundamental concepts on the cloud network slicing context. After that, in Section \ref{sec:protocol}, we detail the research protocol used in the systematic mapping study. In Section \ref{sec:results}, we present the obtained results. An insight of still-open challenges and future directions are discussed in Section \ref{sec:OCFD}.
Section \ref{sec:RQaA} summarizes the research questions and answers. 
Lastly, we present our concluding remarks in Section \ref{sec:conclusions}.

\section{Background and fundamental concepts} \label{sec:background}

We do not intend to make a review of concepts related to slice, since there are dozens of papers about this. However, in order to conceive this work minimally self-contained, we aim to describe the most important aspects related to slice, in line with the main SDOs.

The paradigm shift created by the SDN concept (\citet{Mckeown:2009}) in 2009 opened a new range of options for the operation and management of computer networks. In 2012, the concept of NFV was defined \cite{ETSI:2012} and allowed to perform virtualized network functions on general purpose hardware. Both, SDN and NFV, are enabling technologies that use techniques of network programmability to provide greater flexibility in the management and operation of a network.

The integration between SDN and NFV paved the way for researchers to better exploit the infrastructure resources through an abstraction layer that hides all the network complexity. In this context, several workgroups were created by SDOs to define the fundamental concepts about network slicing. Below, we present some network slicing definitions highlighted by the major SDOs.

\subsection{ETSI}

According to the ETSI, network slicing is a concept that allows the support of logical networks tailored for a specific service, or set of services, over a shared common network infrastructure, for the purpose of efficient utilization of network resources \cite{ETSI:2018}. Based on this concept, the ETSI has defined an architecture for the creation and management of network slices. This architecture allows different network providers to offer slices to concurrent tenants that have different services and requirements. In short, the architecture has three well-defined layers: (1) service instance layer; (2) network instance layer; and (3) resource layer.

\begin{figure}[ht]
    \centering
    \includegraphics[scale=0.43]{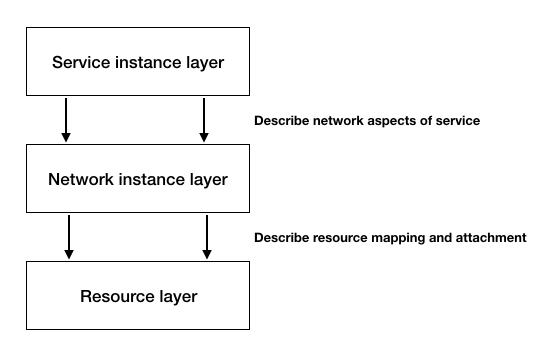}
    \caption{3-layer ETSI architecture. Adapted from \citet{ETSI:2018}.}
    \label{fig:ETSI}
\end{figure}

In Fig. \ref{fig:ETSI}, the service instance layer corresponds to a high-level description of the service. The network instance layer is responsible for abstracting the resource mapping. The resource layer represents physical or virtual devices that could be an offer to a specific service into a slice.

\subsection{IETF}

The IETF has created a working group\footnote{https://datatracker.ietf.org/wg/netslicing/about/.} in the network slicing area. Although no official document has been finalized and no RFC has been published, some drafts have been produced.

In \citet{Galis:2017}, the network slice is treated as managed partitions of physical and/or virtual network resources, network physical/virtual and service functions that can act as an independent instance of a connectivity network and/or as a network cloud. In other draft \cite{Geng:2017}, network slice is defined as a managed group of subsets of resources, network functions / network virtual functions at the data, control, management/orchestration planes and services at a given time. \citet{Wang:2018} describe that the mechanism of network slicing is defined to divide common physical network infrastructure into diverse isolated virtual network resources, to meet the high-level demands from different vertical industries.

\subsection{NGMN}

\begin{figure}[ht]
    \centering
    \includegraphics[scale=0.25]{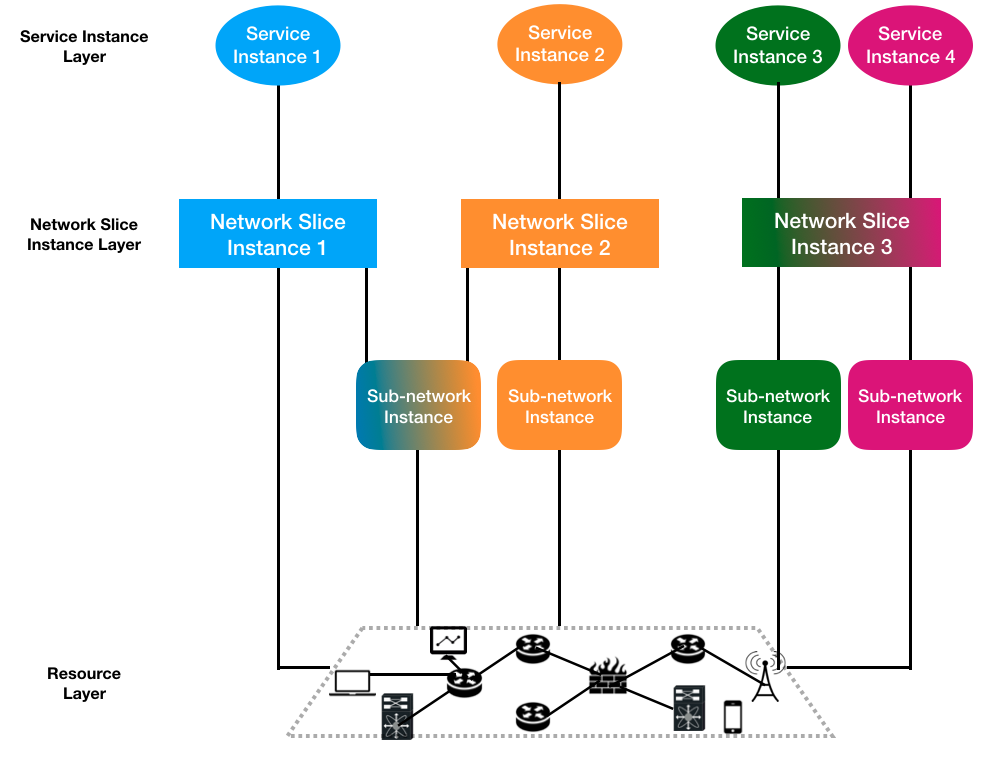}
    \caption{NGMN architecture. Adapted from \citet{NGMN:2016}.}
    \label{fig:NGMN}
\end{figure}

Like ETSI, NGMN \cite{NGMN:2016} defines the network slicing concept in 3 layers: (1) service instance; (2) network slice instance; and (3) resource layer. A concept about sub-network instance is also presented. In this context, various sub-network instances could be used to compose a major network slice instance. A sub-network instance approach is interesting because of its concept of inheritance, which brings the possibility to be shared by another network slice instance. In Fig. \ref{fig:NGMN}, we show the NGMN architecture.

\subsection{3GPP}

\citet{3GPP:2019} defines network slicing as the logical network that provides specific capabilities and characteristics. In 5G context, it is defined as an end-to-end logical communication network, within a Public Land Mobile Network (PLMN). This network is formed by: a Core Network (CN), an User Plane and a 5G Access Network (AN). The concept of a network slice instance (NSI) was created by 3GPP as a managed entity in the operator's network, which has an independent lifecycle compared to the service instance(s) \cite{3GPP:2018}. According to 3GPP, the lifecycle has the following phases: (1) Preparation; (2) Instantiation, configuration and  activation; (3) Run-time; and (4) Decommission.

\subsection{ITU-T}
 
According to ITU-T \cite{ITU-T:2017}, network slice enables the creation of customized networks, called logically isolated network partitions (LINPs), to provide flexible solutions for different market scenarios that have diverse requirements, with respect to functionalities, performance and resource allocation. In this case, physical resources (\textit{routers, switch, hosts, etc.}) are shared among LINPs, that represents a specific service offered by a virtual network. In fact, each LINP is managed by individual LINP managers.

\subsection{(Cloud) Network Slicing}

After shortly describing the background in network slicing, we see that the fundamental concept has a diverse scope. Some definitions, like 3GPP \cite{3GPP:2019} and NGMN \cite{NGMN:2016}, are focused on 5G communications. On the other hand, the definition of ETSI \cite{ETSI:2018} is focused on the description of a service based architecture. The definition of ITU-T \cite{ITU-T:2017} focuses on virtualizing network elements such as routers and switches. The IETF does not have an official definition yet, but the drafts point toward information models of architecture, placing the concept of cloud computing in the context of network slicing.

\begin{figure}[ht]
    \centering
   \frame{\includegraphics[scale=0.25]{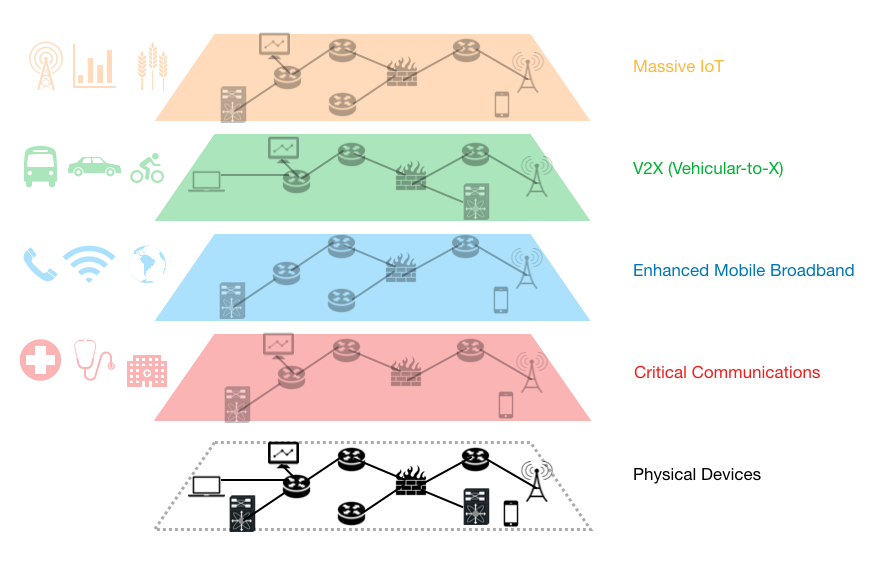}}
    \caption{Network Slicing.}
    \label{fig:Net_slice}
\end{figure}

In this sense, we believe that a better clarification is needed in understanding the concepts presented. First, we must understand that network slicing is being used by the scientific community to define a smarter way to use resources, in order to enable the execution of new (vertical) services on the same shared infrastructure. Second, the resources being shared include networking, cloud, storage, and computing.

For this reason, we conjecture that we should use two concepts: \textbf{Network Slicing} and \textbf{Cloud Network Slicing}. Network slicing is a managed subset of resources, network functions, control, management/orchestration, and service plans at any time \cite{Galis:2018}. In Figure \ref{fig:Net_slice} we can see a network slicing example with different verticals.

\begin{figure}[ht]
    \centering
   \frame{\includegraphics[scale=0.25]{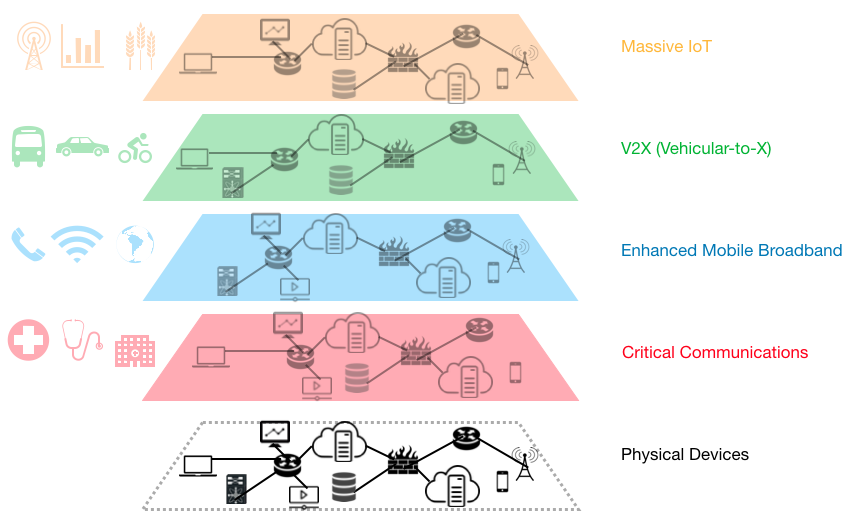}}
    \caption{Cloud Network Slicing.}
    \label{fig:Cloud_Net_slice}
\end{figure}

The NECOS (\textit{Novel Enablers for Cloud Slicing})\footnote{http://www.h2020-necos.eu/.} project coined the term Cloud Network Slicing (CNS) as a set of infrastructures (network, cloud, data center) components/network functions, infrastructure resources (i.e., connectivity, compute, and storage manageable resources) and service functions that have attributes specifically designed to meet the needs of a vertical industry or a service \cite{Silva:2018}. In Fig. \ref{fig:Cloud_Net_slice}, cloud and network elements are shared between different slices, which represent a more up-to-date view of slicing.

In this context, a CNS may consist of cloud and network elements in multi-domain, crossing multiple providers to form an end-to-end slice.

\section{Research protocol} \label{sec:protocol}

During scientific research, researchers usually perform a literature review about a specific topic inside a research area. This step can be performed by means of a systematic review (SR), that goes through existing primary reports, reviews them in-depth and describes their methodology and results. However, applying a SR also has several drawbacks, the main one being that it requires considerable effort \cite{Petersen:2008}. 

On the other hand, a systematic mapping study provides an overview of a research area, identifying the amount, types of search and results available \cite{Petersen:2008}. In this work, we use a systematic mapping study based on \citet{PETERSEN20151} applied to the Cloud Network Slicing context. We adopt a research protocol to guide the stages in this mapping as follows: (1) define the research area; (2) define the research questions; (3) define the search strategy; (4) define exclusion criteria; (5) define the classification process; (6) extract the data and plot a mapping.

\subsection{Research area} 

In the literature, there are systematic mapping studies focused on cloud computing \cite{ABDELMABOUD2015159, NOVAIS2019296, Durao2014}. However,  there is no such a type of study related to network slicing area. In this sense, a study that makes a systematic mapping including the two areas together is highly desirable. 

In our study, we define the scope in the context of \textbf{Cloud Network Slicing}, whose objective is to understand the development of this new area, structuring and categorizing scientific research that were published in the last 10 years.

\subsection{Research questions} \label{sec:RQ}

Research questions set a direction for the mapping study so that the frequencies of publications over time can be mapped and trends can be highlighted \cite{Petersen:2008}. In this study, the following research questions were defined:

\begin{enumerate}
    \item \textbf{RQ1:} What are the main companies that make research on CNS?
    \item \textbf{RQ2:} What are the most cited papers in CNS?
    \item \textbf{RQ3:} Who are the most cited researchers in CNS?
    \item \textbf{RQ4:} How many publications about CNS have been published in the last 10 years?
    \item \textbf{RQ5:} What are the top places used so far for publishing papers on CNS?
    \item \textbf{RQ6:} What are the most developed areas in CNS?
    \item \textbf{RQ7:} Is it possible to classify papers according to a taxonomy? If so, what would it be?
    \item \textbf{RQ8:} What are the most frequently applied research methods, and in what study context?
     \item \textbf{RQ9:} What are the open challenges in CNS?
\end{enumerate}

The answers to these research questions will contribute to give a step forward regarding this area since it makes it possible to understand the direction of existing research and why there are areas not yet researched. We later answer these questions after showing the obtained results.

\subsection{Search strategy}

For the sake of searching, we define a generic string that was submitted to a group of search engines for searching relevant papers. In this study, the generic string used was:

\begin{framed}
\noindent{\fontfamily{pcr}\selectfont
\scriptsize\centering ((((Network) OR (Cloud)) AND (Slicing)) AND (Management))\\
OR\\  
((((Network) OR (Cloud)) AND (Slicing)) AND (Orchestration))\\ 
OR\\ 
((((Network) OR (Cloud)) AND (Slicing)) AND (Intent-Based Network))\\ 
OR\\ 
((((Network) OR (Cloud)) AND (Slicing)) AND (Artificial Intelligence))\\ 
OR\\ 
((((Network) OR (Cloud)) AND (Slicing)) AND 
(Service Assurance))\\ 
OR\\ 
((((Network) OR (Cloud)) AND (Slicing)) AND (Elasticity))\\ 
OR\\ 
((((Network) OR (Cloud)) AND (Slicing)) AND (5G))\\ 
OR\\ 
((((Network) OR (Cloud)) AND (Slicing)) AND (Pricing))\\ 
OR\\ 
((((Network) OR (Cloud)) AND (Slicing)) AND (Architecture))\\} 
\end{framed}

We submitted this search string to the main search engines, as listed below:

\begin{itemize}
    \item IEEE Xplore (Types: Conferences, Journals and Magazines [2009-2019]);
    \item ACM Digital Library (Types: Proceeding and Periodical [2009-2019]);
    \item Science Direct (Elsevier)  (Types: Research and Review articles; [2009-2019])
    \begin{itemize}
        \item Computer Communications;
        \item Computer Networks;
         \item Journal of Network and Computer Applications.
    \end{itemize}
    \item Springer (Types: Research and Review articles [2009-2019]);
    \begin{itemize}
        \item JISA (Journal of Internet Services and Applications);
        \item JNSM (Journal of Network and Systems Management).
    \end{itemize}
\end{itemize}

The query was performed on January 14th, 2020 and \textbf{1696} indexed papers matched the search. All indexed studies were downloaded for analysis.

\subsection{Exclusion criteria}

Exclusion criteria were defined to remove studies that are unrelated to the research objective. They are listed below:

\begin{itemize}
    \item \textbf{EC1:} Papers with irrelevant content to the search area;
    \item \textbf{EC2:} Duplicated papers;
    \item \textbf{EC3:} Papers that deal with "Management" but are not related to Cloud Network Slicing;
    \item \textbf{EC4:} Papers that deal with "Orchestration" but are not related to Cloud Network Slicing;
    \item \textbf{EC5:} Papers that deal with "5G" but are not related to Cloud Network Slicing;
    \item \textbf{EC6:} Papers that deal with "Pricing" but are not related to Cloud Network Slicing;
    \item \textbf{EC7:} Papers that deal with "Architecture" but are not related to Cloud Network Slicing.
    \item \textbf{EC8:} Studies that are not full papers (short papers, demos and posters).
\end{itemize}

After this stage, \textbf{640} studies were included for an in-depth analysis.

\subsection{Classification process} \label{sec:Class_proc}

In this stage, the focus was to quickly read and classify all the 640 papers. In our study, the classification process analyzes the correlation between two facets: \textbf{research facet (RF)} and \textbf{technological facet (TF)}. These facets served as the basis for the definition of a new taxonomy in the CNS context.

The research facets were defined based on the classification process proposed by \citet{Wieringa:2005}, as follow:

\begin{itemize}
    \item \textbf{RF1 - Evaluation Research:} techniques are implemented in practice and an evaluation of the technique is conducted. This type of papers show how the technique is implemented in practice (solution implementation) and what are the consequences of the implementation in terms of benefits and drawbacks (implementation evaluation). This also includes identifying problems in the industry.
    \item \textbf{RF2 - Solution Proposal:} a solution for a problem is proposed. The solution can be either novel or a significant extension of an existing technique. The potential benefits and the applicability of the solution are shown by a small example or a good line of argumentation.
    \item \textbf{RF3 - Validation Research:} techniques investigated are novel and have not yet been implemented in practice. Techniques used are, for example, experiments, i.e., work done in the lab.
    \item \textbf{RF4 - Philosophical papers:} these papers sketch a new way of looking at existing things by structuring the field in the form of taxonomy or conceptual framework.
    \item \textbf{RF5 - Personal experience papers:} experience papers explain what and how something has been done in practice. It has to be the personal experience of the author.
    \item \textbf{RF6 - Opinion papers:} these papers express the personal opinion of somebody whether a certain technique is good or bad, or how things should be done. They do not rely on related work and research methodologies.
\end{itemize}

The technological facets were defined by analyzing the frequency of keywords in the indexed articles. Terms with similar meanings have been grouped together for a more objective classification process. For example, papers that address artificial intelligence to make automated decisions have been placed inside the orchestration facet. In a nutshell, technological facet defines the scope of this study in the CNS context. 

That said, the technological facets used were divided into five categories, as follows:

\begin{itemize}
     \item \textbf{TF1 - Pricing model}: Fixed, Dynamic or Mixed.
    \item \textbf{TF2 - Orchestration}: Artificial Intelligence, Intent-based Network, Service Assurance or Elasticity;
    \item \textbf{TF3 - 5G}: RAN (Radio Access Network), Transport Network or Core Network;
    \item \textbf{TF4 - Architecture}: Single-domain or Multi-domain;
    \item \textbf{TF5 - Management}: Fault, Configuration, Accounting, Performance and Security.
\end{itemize}

After this stage, data were collected and stored in a database to look for evidence from scientific publications over time in the context of CNS.

\section{Results} \label{sec:results}

In this section, we present the quantitative results obtained with the systematic mapping study in the field of CNS. The results presented here have a direct relationship with the research questions defined in Subsection \ref{sec:RQ}. In fact, from the obtained results we are able to answer the nine elaborated questions.

\subsection{Industry involvement}

One of the first findings is related to the industry participation in the indexed papers. Table \ref{tab:Industry} presents the top 10 companies and answers \textbf{RQ1}.

\begin{table}[ht] \footnotesize
\renewcommand{\arraystretch}{1.5}
    \centering
    \begin{tabular}{c c}
        \hline
        Company & Participations \\
        \hline
         Nokia Bell Labs & 57  \\
         Huawei & 51  \\
         NEC & 38  \\
         Ericsson & 26  \\
         Deustche Telekom & 13  \\
         Telecom & 12  \\
         Telefónica & 12  \\
         Nextworks & 11  \\
         IMDEA & 9  \\
         Samsung & 9  \\
        \hline
        \\
    \end{tabular}
\caption{Top 10 companies.}
\label{tab:Industry}
\end{table}

We note that Nokia \cite{NOKIA:2019}, Huawei \cite{HUAWEI:2019}, NEC \cite{NEC:2019} and Ericsson \cite{ERICSSON:2019} already have products or prototypes in the network slicing area. Nokia has a product focused on slicing the access network called \textit{Nokia Fixed Access Network Slicing}. Huawei has a solution called \textit{eLTE-DA Smart Grid Solution} using slicing, aimed at the smart grid industry. NEC, in conjunction with Netcracker Inc., launched a management solution called \textit{HOM (Hybrid Operations Management)} based on slicing. Ericsson, together with SK Telecom, demonstrated success in the execution of 5G slicing prototypes.

Overall, companies are expected to invest in scientific research to earn a return (ROI - Return on Investment) on products or services, although we observed that just 39\% of the studies have some involvement with the industry. In this context, knowing the major companies that have some relation to scientific research in an area can be a key aspect for researchers seeking investments and partnerships.

\subsection{Most cited papers}

In response to \textbf{RQ2}, Table \ref{tab:Most_cited_papers} shows the most cited papers in the CNS area. It is expected that older papers are more likely to have a greater number of citations. In addition, the number of citations is dynamic, i.e., the data presented here are the ones obtained in the day we ran the search query. For this reason, our intention here is not to create a rank, but rather to help direct future research in the area of CNS. Next, we observe some aspects of these papers.

\begin{table}[ht] \footnotesize
\renewcommand{\arraystretch}{1.5}
    \centering
    \begin{tabular}{p{6cm} c c}
        \hline
        Title & Citations & Year\\
        \hline
         Resource management for Infrastructure as a Service (IaaS) in cloud computing: A survey & 256 & 2013\\
         NVS: A Substrate for Virtualizing Wireless Resources in Cellular Networks & 194 & 2012\\
         5G roadmap: 10 key enabling technologies & 158 & 2016 \\
         From Network Sharing to Multi-Tenancy: The 5G Network Slice Broker & 145 & 2016 \\
         Resource Slicing in Virtual Wireless Networks: A Survey & 128 & 2016 \\
         Network Slicing in 5G: Survey and Challenges & 127 & 2017 \\
         Network Slicing for 5G with SDN/NFV: Concepts, Architectures, and Challenges & 125 & 2017 \\
         Information-Centric Network Function Virtualization over 5G Mobile Wireless Networks & 121 & 2015 \\
         Mobile Network Architecture Evolution toward 5G & 109 & 2016 \\
         Network Slicing Based 5G and Future Mobile Networks: Mobility, Resource Management, and Challenges & 101 & 2017 \\
        \hline
        \\
    \end{tabular}
\caption{Most cited papers.}
\label{tab:Most_cited_papers}
\end{table}

The paper entitled \textit{``Resource management for Infrastructure as a Service (IaaS) in cloud computing: A survey''} was published in 2014 and has 256 citations. This survey focuses on resource management techniques that tackle problems such as resource provisioning, resource allocation, resource mapping and resource adaptation. In addition, open challenges in resource management are pointed out.

The second most cited paper, with 194 citations, was \textit{``NVS: A Substrate for Virtualizing Wireless Resources in Cellular Networks''}. This study proposes the design and implementation of a network virtualization substrate for the effective virtualization of wireless resources in cellular networks. In a nutshell, this paper brings a way to run slices simultaneously with different types of reservations.

\textit{``5G roadmap: 10 key enabling technologies''} has obtained 158 citations and presents the state-of-the-art of ten potential technologies in 5G environments, such as: 1) wireless software-defined network, (2) network function virtualization, (3) millimeter wave spectrum, (4) massive MIMO, (5) network ultra-densification, (6) big data and mobile cloud computing, (7) scalable Internet of Things, (8) device-to-device connectivity with high mobility, (9) green communications, and (10) new radio access techniques.

With 145 citations, the paper \textit{``From Network Sharing to Multi-Tenancy: The 5G Network Slice Broker''} presents an overview of the 3GPP standard evolution; from network sharing principles, mechanisms, and architectures to future on-demand multi-tenant systems, focusing on the concept of the 5G Network Slice Broker.

The paper \textit{``Resource Slicing in Virtual Wireless Networks: A Survey''}, with 128 citations, is a study that focuses on isolation issues in slicing environment. It discusses how technologies such as SDN and NFV can help with resource slicing solutions. 

\textit{``Network Slicing in 5G: Survey and Challenges''}, with 127 citations, is a paper that brings a review of the state-of-art in 5G network slicing and presents a framework to evaluate the maturity of current proposals and identify open research issues.

The paper \textit{``Network Slicing for 5G with SDN/NFV: Concepts, Architectures, and Challenges''} has obtained 125 citations and brings a study of network slicing focused in 5G environment. In this paper, SDN and NFV capabilities were analyzed from ETSI perspectives.

With 121 citations, the paper \textit{``Information-Centric Network Function Virtualization over 5G Mobile Wireless Networks''} presents a way to integrate wireless network virtualization and information-centric networking techniques. The authors formulate a virtual resource allocation and in-network caching strategy for architecture optimization. 

\textit{``Mobile Network Architecture Evolution toward 5G''}, with 109 citations, discusses 3GPP mobile network evolution focusing on some key topics, such as: network functions virtualization, network slicing, software-defined mobile network control, management, and orchestration.

The paper \textit{``Network Slicing Based 5G and Future Mobile Networks: Mobility, Resource Management, and Challenges''}, with 101 citations, presents a scheme for managing mobility among different access networks. In addition, open issues and challenges in network-slicing-based 5G networks are discussed, including network reconstruction, network slicing management, and cooperation with other 5G technologies.

In the mapping study realized by \citet{Abdelmaboud:2015}, the authors performed a structural analysis in the 67 indexed papers obtained in their search. In that case, for each paper, they collected details such as problem addressed, basic approach, scope, limitation approach, validation, and the result of the validation.

The \textbf{Problem addressed} is a brief description of what is covered in the paper. The \textbf{Basic approach}  makes reference to the type of contribution described in the paper. The \textbf{Scope} is a brief description about the focus of the article. \textbf{Limitations} are related to issues not solved in the study. \textbf{Validation} represents whether the study performed any type of experiment to prove (validate) the research hypothesis. The \textbf{Result} highlights the achievements of the paper. 

However, in our study, a total of 640 papers were indexed and as such, dDoing a structural analysis in these articles, as proposed in \citet{Abdelmaboud:2015}, would require a lot of effort and time. For this reason, we performed the structural analysis for the 10 most cited papers in Table \ref{tab:Most_cited_papers}, and the result of this analysis is presented in \ref{ap:StrucAna}.

\subsection{Most active researchers}

The top researchers in a particular research area become references, and their research can define the future directions of that specific area. We answer \textbf{RQ3} by listing in Table \ref{tab:Researches} the Top 10 most-cited researchers in the field of CNS.

\begin{table}[ht] 
\renewcommand{\arraystretch}{1.5}
        \scalefont{0.9}
    \centering
    \begin{tabular}{c l c}
        \hline
        Name & Organization & Quantify\\
        \hline
         Xavier Costa-Pérez & NEC Laboratories Europa & 21  \\
         Navid Nikaein & Eurecom & 15 \\
         Gang Feng & Uni. of Elec. Sc. and Tech. of China & 14  \\
         Vincenzo Sciancalepore & NEC Laboratories Europa & 14  \\
         Albert Banchs & Univ. Carlos III of Madrid & 13  \\
         Tarik Taleb & Aalto University & 12  \\
         Konstantinos Samdanis & Nokia Bell Labs & 11  \\
         Jordi Pérez-Romero & Universidad Politécnica de Cataluña & 11  \\
         Adlen Ksentini & Eurecom & 10  \\
         Shuang Qin & Uni. of Elec. Sc. and Tech. of China & 10  \\
        \hline
        \\
    \end{tabular}
\caption{Top 10 Researchers.}
\label{tab:Researches}
\end{table}

We conjecture that knowing the most influential researchers in a research area can help students focus their studies and fostering future partnerships with other researchers. In the industry, the participation of an influential researcher may be a key factor in choosing a project or partnership. For this reason, we present below a brief summary of each one's activities.

Xavier Costa-Pérez has M.Sc. and Ph.D. degrees in telecommunications from the Polytechnic University of Catalonia (UPC) in Barcelona. He is the head of 5G networks R\&D and deputy general manager of the security \& networking R\&D division at NEC laboratories Europe.

Navid Nikaein is professor in the communication systems department at Eurecom, where he is leading a R\&D group on experimental system research related to 4G-5G wireless systems and networking protocols, as well as agile service delivery platforms. He has a Ph.D degree in communication systems from the Swiss Federal Institute of Technology (EPFL).

Gang Feng has a Master degree in Electronic Engineering from the University of Electronic Science and Technology of China (UESTC) and a Ph.D. degree in Information Engineering from The Chinese University of Hong Kong. Currently, he is a professor with the National Laboratory of Communications, University of Electronic Science and Technology of China.

Vicenzo Sciancalepore has a M.Sc. degree in telecommunications engineering and telematics engineering, and he received a double degree Ph.D. from Politecnico di Milano and Universidad Carlos III de Madrid. He is a senior researcher and RAN specialist at NEC laboratories Europe GmbH, Germany.

Albert Banchs received his telecommunications engineering and Ph.D. degrees from UPC BarcelonaTech, Spain. Since 2009, he also has a double affiliation as Deputy Director of the IMDEA Networks research institute and Professor at University Carlos III of Madrid.

Tarik Taleb is associate professor from Aalto University, Finland. He has a Master's and Ph.D. degrees in information sciences from Tohoku University.

Konstantinos Samdanis is a research project manager at Nokia Bell Labs in Germany. He has a Master of Science and Ph.D. degrees in mobile communications from King's College London.

Jordi Pérez-Romero is a researcher from the Polytechnic University of Catalonia (UPC). He has telecommunications engineering degree and the Ph.D. from Polytechnic University of Catalonia (UPC).

Adlen Ksentini is a professor from Eurecom and has a master's degree in networking and multimedia at the University of Versailles and a Ph.D. in computer science from University of Cergy-Pontoise.

Shuang Qin has a B.E. degree in electronic information science and technology and Ph.D. degree in communication and information system from UESTC. He is currently an Associate Professor with the National Key Laboratory of Science and Technology on Communications, UESTC.

In what regards the top 10 researchers, we note that eight in the top ten are from Europe. Organizations such as ETSI may explain the strong involvement of researchers from Europe in the context of CNS. This probably indicates that the major developments and directions in the area of CNS have received greater focus in Europe.

\subsection{Number of publications over the years}

Knowing the history of scientific publications in a research area allows us to observe whether the topic is still interesting to the scientific community. Table \ref{tab:Freq_pub} presents the number of publications over the years in the CNS context and answers \textbf{RQ4}.

\begin{table}[ht]
\renewcommand{\arraystretch}{1.5}
    \scalefont{0.78}
    \centering
    \begin{tabular}{ c c c c c c c c c c}
        \hline
        2010 & 2011 & 2012 & 2013 & 2014 & 2015 & 2016 & 2017 & 2018 & 2019\\
        \hline
        4 & 4 & 5 & 6 & 13 & 24 & 40 & 106 & 219 & 219 \\
        \hline
        \\
    \end{tabular}
\caption{Frequency of publications over the years.}
\label{tab:Freq_pub}
\end{table}

\begin{figure*}[ht]
\centering
   \subfigure[SDN]{ 
   \includegraphics[scale=0.28]{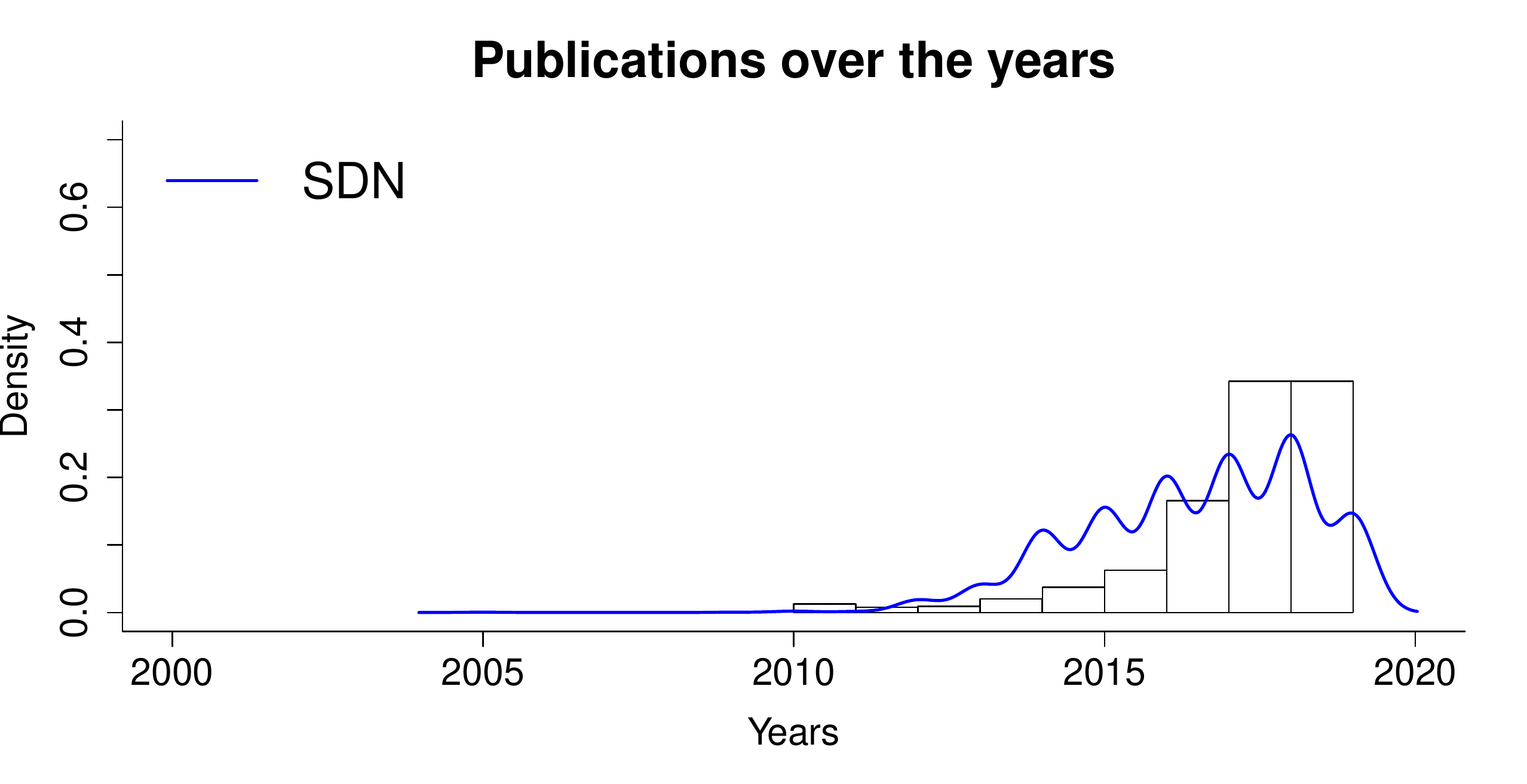}
   \label{fig:Hist-SDN}
   }
   \subfigure[NFV]{
   \includegraphics[scale=0.28]{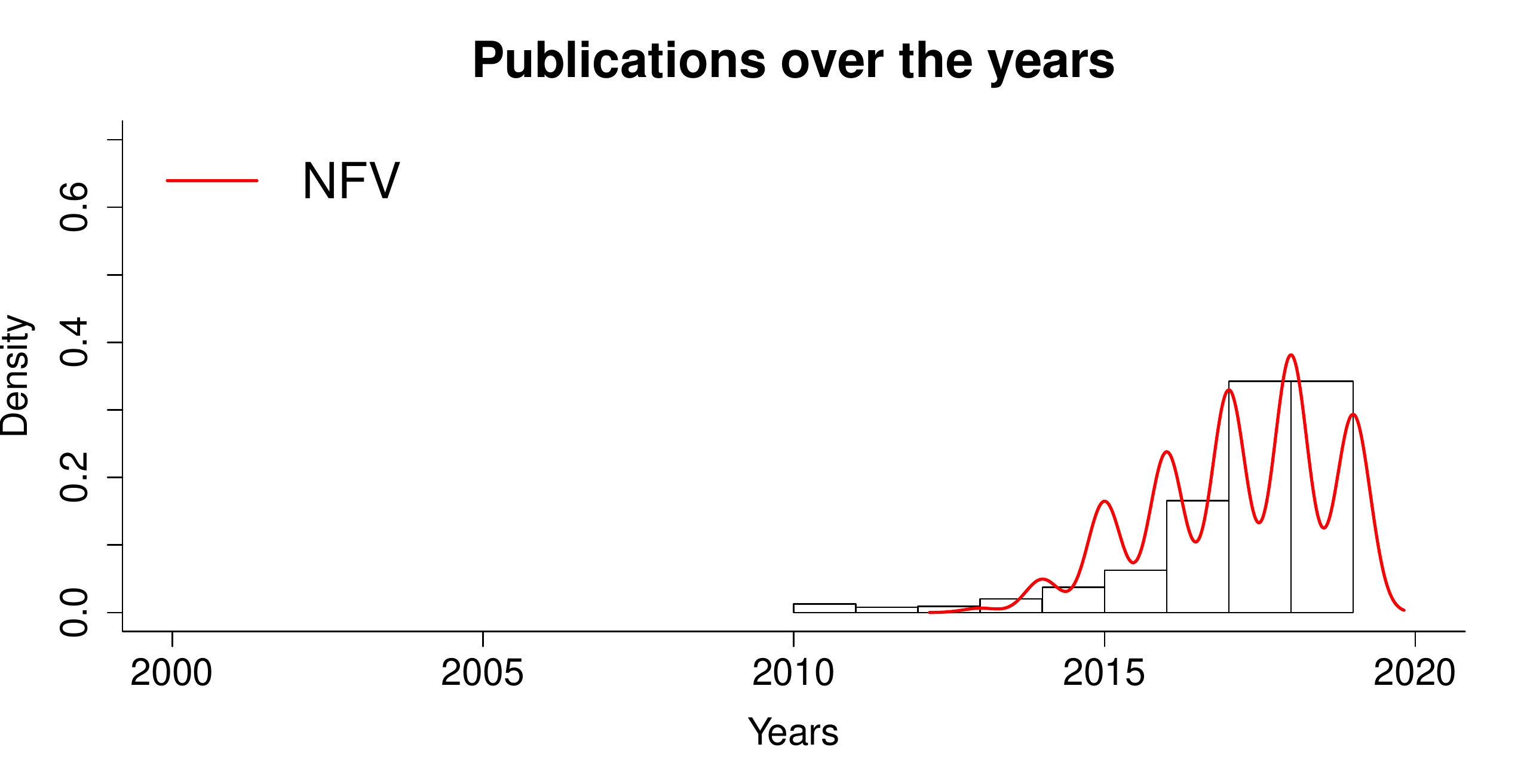}
   \label{fig:Hist-NFV}
   }
   \subfigure[Cloud Computing]{
   \includegraphics[scale=0.28]{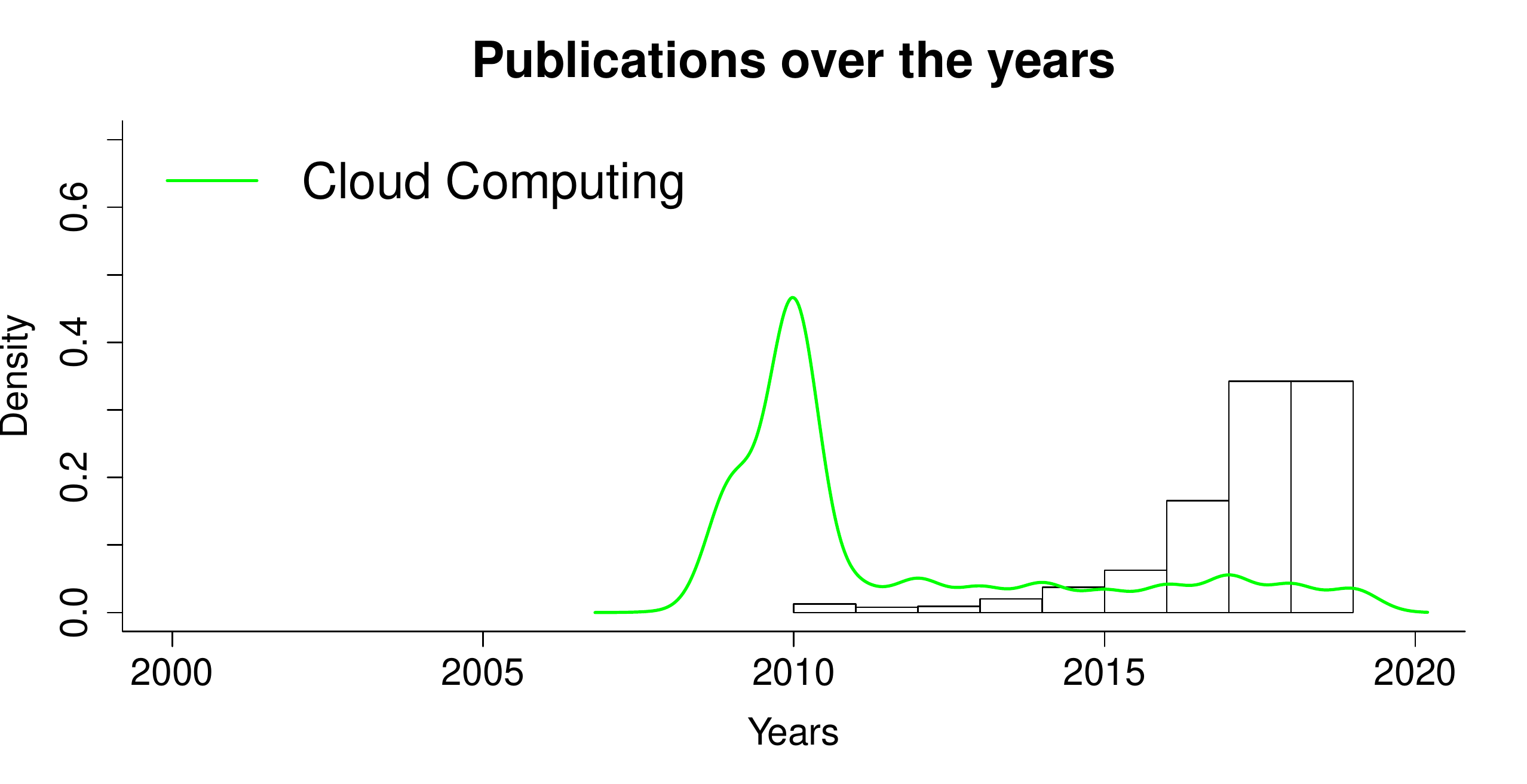}
   \label{fig:Hist-Cloud}
   }
   \subfigure[Virtualization]{
   \includegraphics[scale=0.28]{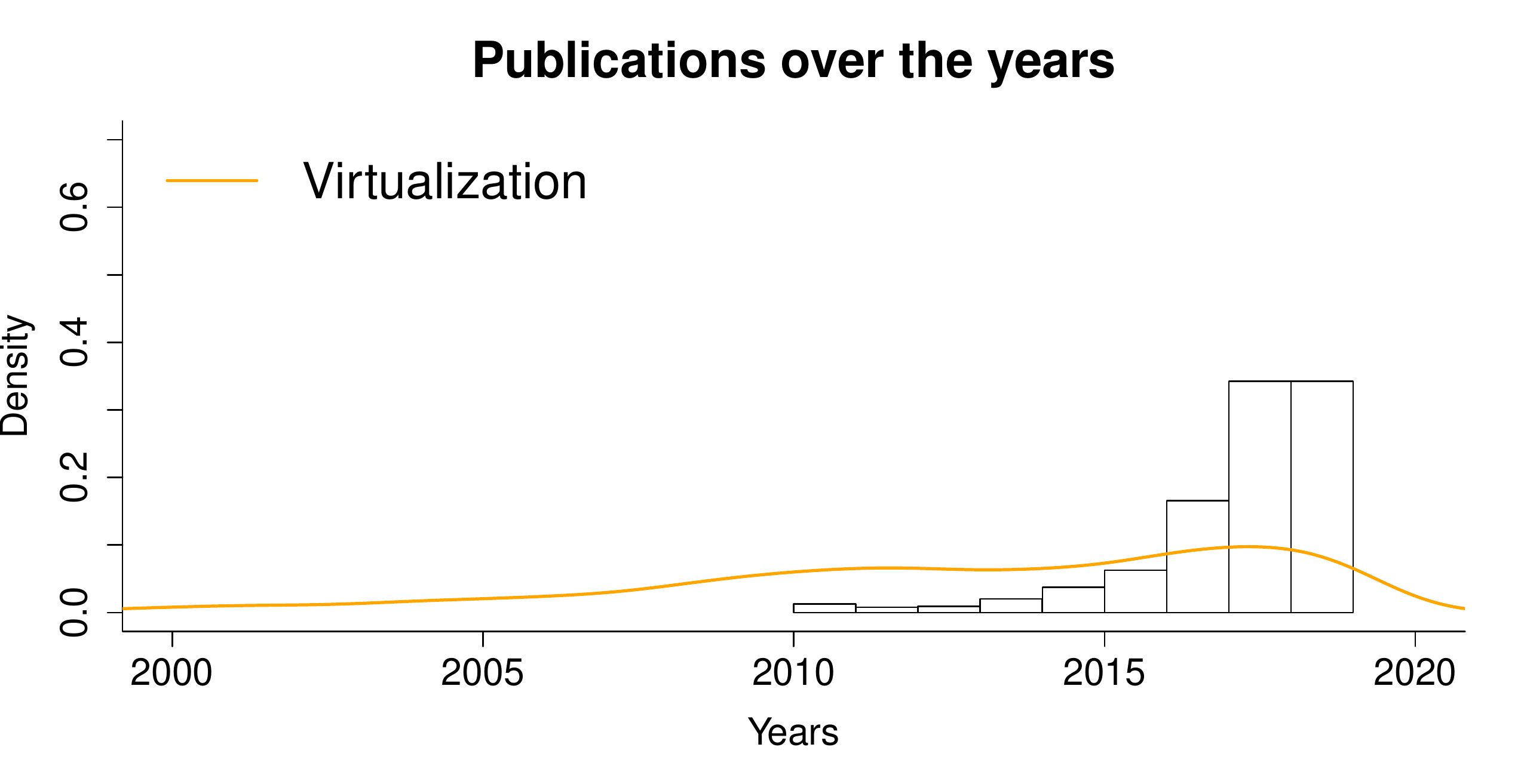}
   \label{fig:Hist-Virt}
   } 
   \caption{Publications over the years (Related Areas).}
   \label{fig:Pub_Years_RA}
\end{figure*}

From Table \ref{tab:Freq_pub}, we see that the number of publications has increased significantly in the last ten years, from four publications in 2010 to 219 in the last two years. 

To improve the understanding about these numbers, we decided to compare the number of publications in areas related to the context of CNS. After analyzing the frequency of keywords from indexed articles, we obtain the following research areas related to the context of the CNS: SDN, NFV, Cloud Computing and Virtualization. In this case, we analyzed the density of papers distributed over the years about CNS regarding the number of publications in the above related areas. We prefer to use the density due to the high variability in the number of publications in related areas over the years.

Figure \ref{fig:Pub_Years_RA} depicts the density of the publications. The histogram represents the number of papers indexed over the years in CNS and the lines represent the number of papers published in similar areas (SDN, NFV, Cloud Computing and Virtualization). Almost 85\% of the publications in CNS are distributed between 2017-2019. The behavior of SDN and NFV curves are similar to the CNS histogram, with peak in 2018. The cloud computing and virtualization curves have different behavior, where the former has a huge peak in 2010 and the latter presents a more flat behavior over the years, with a small peak in 2017.

It's interesting to observe that in related areas, lines have a decrease after a peak. We observe also that the CNS research topic has not yet peaked, considering that the number of publications on this subject has not decreased so far. This may be a good evidence that CNS research still has challenges that have not been fully explored.

We also decided to do a similar analysis with areas not related to CNS. So, we decided to use publications related to DTN (Delay Tolerant Network), Grid Computing and P2P (Peer-to-Peer).

\begin{figure*}[ht]
    \centering
   \subfigure[DTN]{ 
    \includegraphics[scale=0.28]{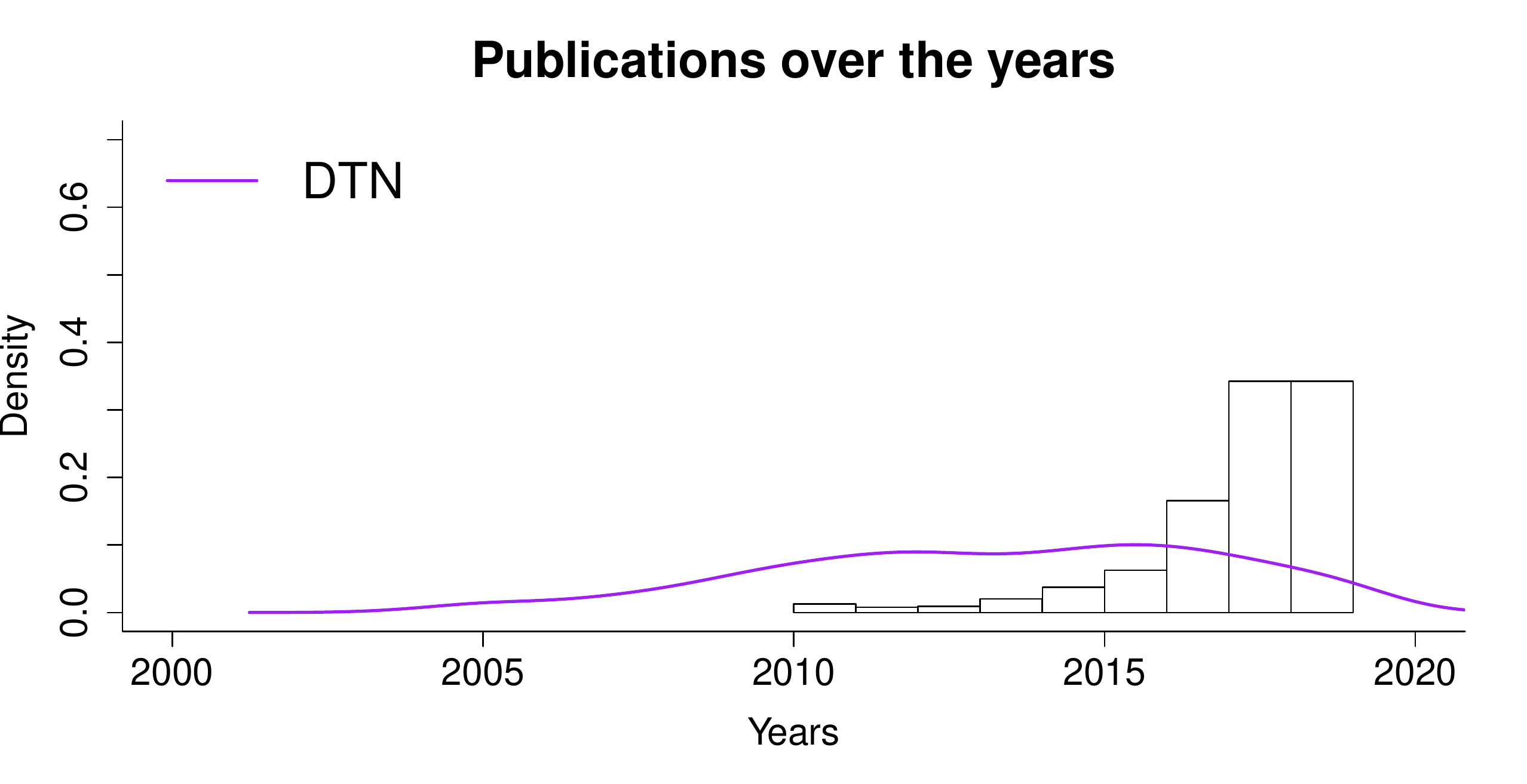}
    \label{fig:Hist-DTN}
   } 
   \subfigure[Grid Computing]{ 
    \includegraphics[scale=0.28]{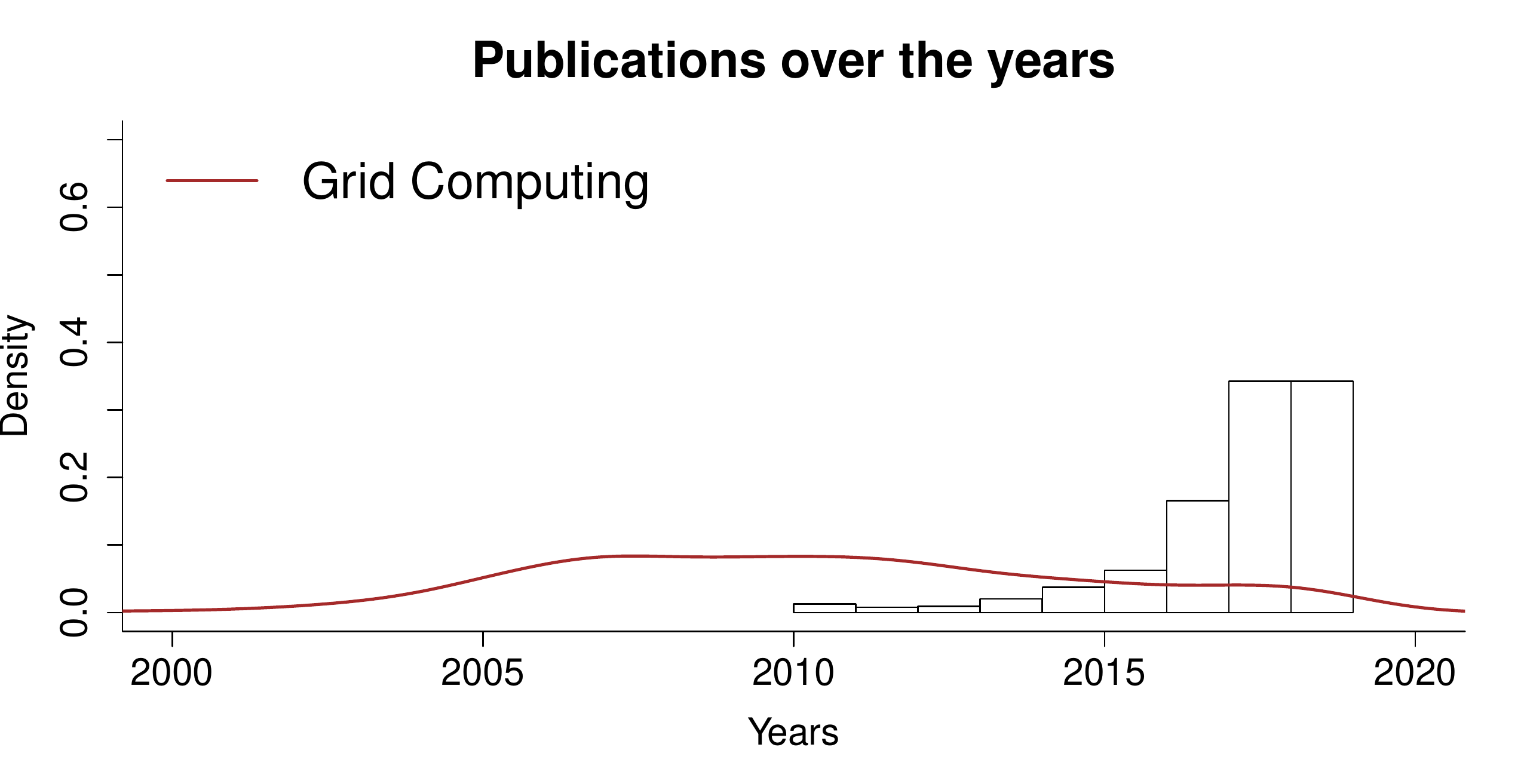}
    \label{fig:Hist-Grid}
   } 
   \begin{center}
    \subfigure[Peer-to-Peer]{
    \includegraphics[scale=0.28]{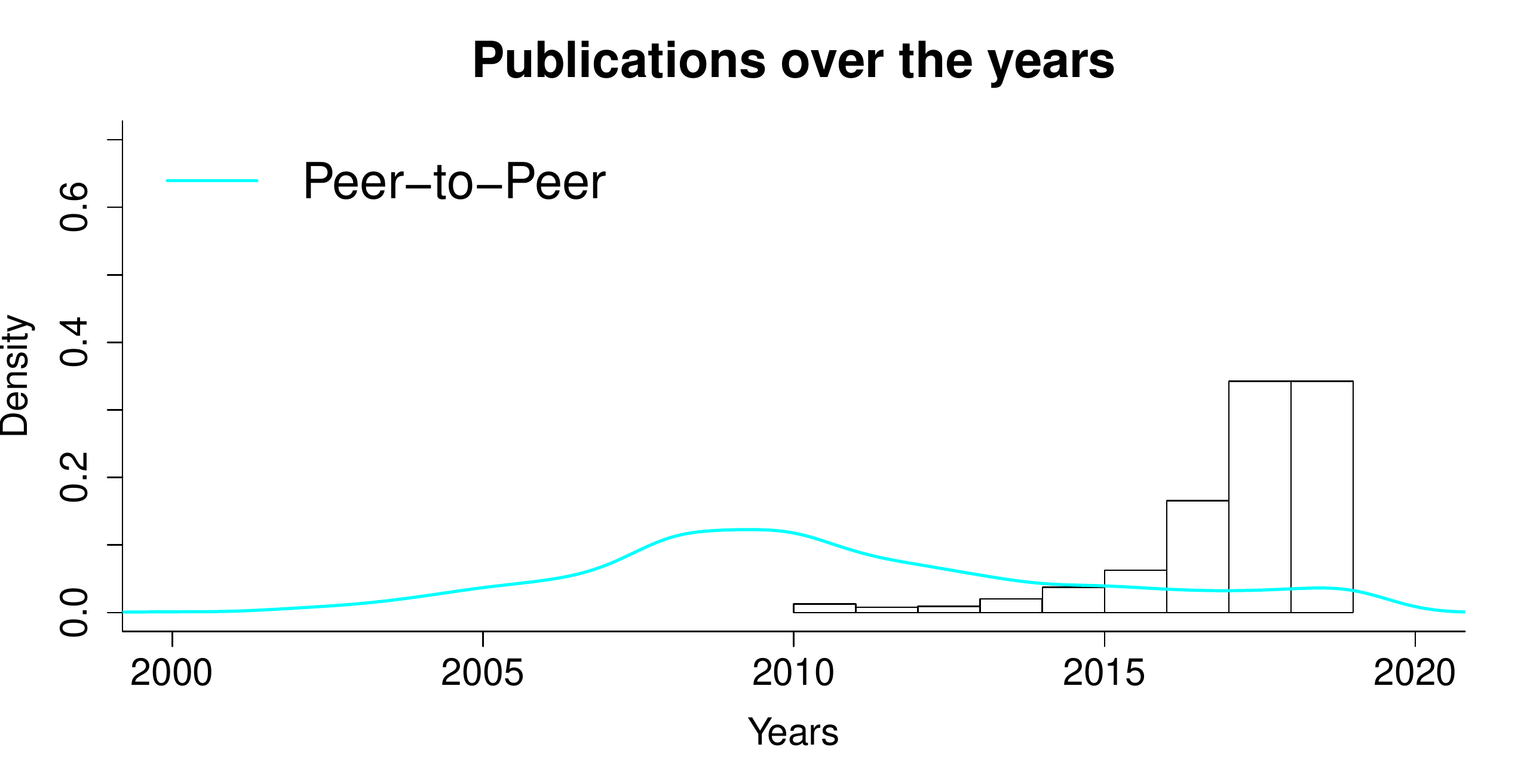}
    \label{fig:Hist-P2P}
   } 
   \end{center}
   \caption{Publications over the years (Unrelated Areas)}
   \label{fig:Pub_Years_URA}
\end{figure*}

Figure \ref{fig:Pub_Years_URA} shows the same distribution (histogram) about CNS, as shown in Fig. \ref{fig:Pub_Years_RA}, but now regarding the number of publications with unrelated areas. In this case, the lines represent the density of papers published in unrelated areas. For the same reason as the related areas, here we use the density due to the high variability in the number of publications in related areas over the years. 

The lines of the three unrelated areas present a moment of growth, reach a peak and then decrease. DTN peaked in 2017, Grid computing peaked in 2007, and finally, P2P had the highest number of publications in 2010. As we know, both grid computing and P2P are no longer considered hot topics by the research community, so it is natural to have a lower number of publications in recent years.

Analyzing the behavior of the curves in Figures \ref{fig:Pub_Years_RA} and \ref{fig:Pub_Years_URA}, we observed that the number of publications from related and unrelated areas to CNS had a peak moment and a decrease in the number of published papers. Based on these data, we cannot say if publications in CNS are already at their peak, but we can say that, as they have not yet decreased, the area still has open challenges for the coming years.

\subsection{Publication venues}

By knowing the main publication venues, researchers can observe other related studies and define strategies about where to submit a new paper. In this context, Table \ref{tab:Places} presents the top ten publication venues in the area of CNS and also answers \textbf{RQ5}.

\begin{table}[ht!] \footnotesize
\renewcommand{\arraystretch}{1.5}
    \centering
    \begin{tabular}{ c c c}
        \hline
        Place & Quantity & Type\\
        \hline
         IEEE Access & 39 & Periodical  \\
         IEEE ICC & 30 & Conference \\
         IEEE Communications Magazine & 26 & Periodical  \\
         EuCNC & 26 & Conference  \\
         IEEE GLOBECOM & 21 & Conference  \\
         IEEE CSCN & 18 & Conference  \\
         JSAC & 18 & Periodical  \\
         IEEE TVT & 16 & Periodical  \\
         IEEE/IFIP NOMS & 13 & Conference  \\
         IEEE WCNC & 13 & Conference  \\
         
        \hline
        \\
    \end{tabular}
\caption{Top 10 publication venues.}
\label{tab:Places}
\end{table}

Table \ref{tab:Places} shows that in the top ten publication venues, four are journals and six are conferences. We believe that research published in journals has a higher maturity level than conferences. This may reinforce the notion that the CNS area still has open challenges as more papers are published in conferences. Next, we briefly summarize the main publication venues in the CNS context.

IEEE Access is a multidisciplinary journal with impact factor 4.098 in 2018 \footnote{According to JCR, available at: https://clarivate.com/webofsciencegroup/solutions\\/journal-citation-reports/.}. It is published in open-access format \cite{OpenAccess:2019}, that is, has unrestricted online access and has no page limits. IEEE Access is indexed by IET Inspec, Ei Compendex, Scopus, EBSCOhost, and Google Scholar. 

IEEE ICC (International Conference on Communications) is an annual conference dedicated to driving innovation in nearly every aspect of communications. The conference program includes technical papers, tutorials, workshops, and industry sessions.

IEEE Communications Magazine is a monthly technical magazine published by the IEEE Communications Society (ComSoc), with a 10.356\footnote{https://ieeexplore.ieee.org/xpl/aboutJournal.jsp?punumber=35\#titleHistory.} impact factor. It focuses on three main topics: (1) communication, networking and broadcast technologies; (2) signal processing and analysis; and (3) computing and processing.

EuCNC (European Conference on Networks and Communications) is a conference sponsored by the IEEE Communications Society and the European Association for Signal Processing. It is supported by the European Commission, focusing on communication networks and systems, reaching services and applications. This conference has oral and poster sessions, panels, tutorials, workshops and keynotes presentations.

IEEE GLOBECOM (Global Communications Conference) is an annual conference organized by the IEEE ComSoc. It has an extensive conference program, including technical panels, demos, tutorials, workshops, and industry presentations.

IEEE CSCN (Conference on Standards for Communications and Networking) is a conference sponsored by IEEE ComSoc and focused on standards-related topics in the areas of communications, networking, cloud computing, and associated disciplines.

Journal on Selected Areas in Communications (JSAC) is a journal that has a focus on communications and networking, with 7.172 impact factor\footnote{https://scijournal.org/impact-factor-of-IEEE-J-SEL-AREA-COMM.shtml.}. This journal uses periodical call for papers with collections in the form of special issues. It is a hybrid journal that permits both traditional subscription-based content, as well as open access (author-paid content).

IEEE TVT (Transactions on Vehicular Technology) is a journal with 5.339\footnote{https://ieeexplore.ieee.org/xpl/aboutJournal.jsp?punumber=25.} impact factor, focused on research regarding the theory and practice of electrical and electronics technology in vehicles and vehicular systems. In this case, it is interesting to observe that a journal focused on vehicular technologies has a lot of publications in the CNS context, which may be an indication that the community behind this research topic is strongly interested in slicing and its benefits. 

IEEE/IFIP NOMS (Network Operations and Management Symposium) is a symposium held every two years (odd ones), organized by IEEE ComSoc and IFIP (International Federation for Information Processing). It has a program including keynotes, panels, technical sessions, demo sessions, dissertation sessions, mini-conference sessions, poster sessions, tutorials, and workshops.

IEEE WCNC (Wireless Communications and Networking Conference) is focused on the advancement of wireless communications and networking technology. The conference program includes workshops, tutorials, keynote talks from industrial leaders, and panel discussions.

\subsection{Holistic view}

A scientific area is usually composed by several sub-areas in which specific research problems are addressed. Knowing the level of developments in these sub-areas can help researchers both understand and direct them. In this context, this subsection presents a holistic view of scientific publications in the CNS area and answers \textbf{RQs\{6,7,8\}}.

\begin{figure*}[hbt!]
    \centering
    \includegraphics[scale=0.50]{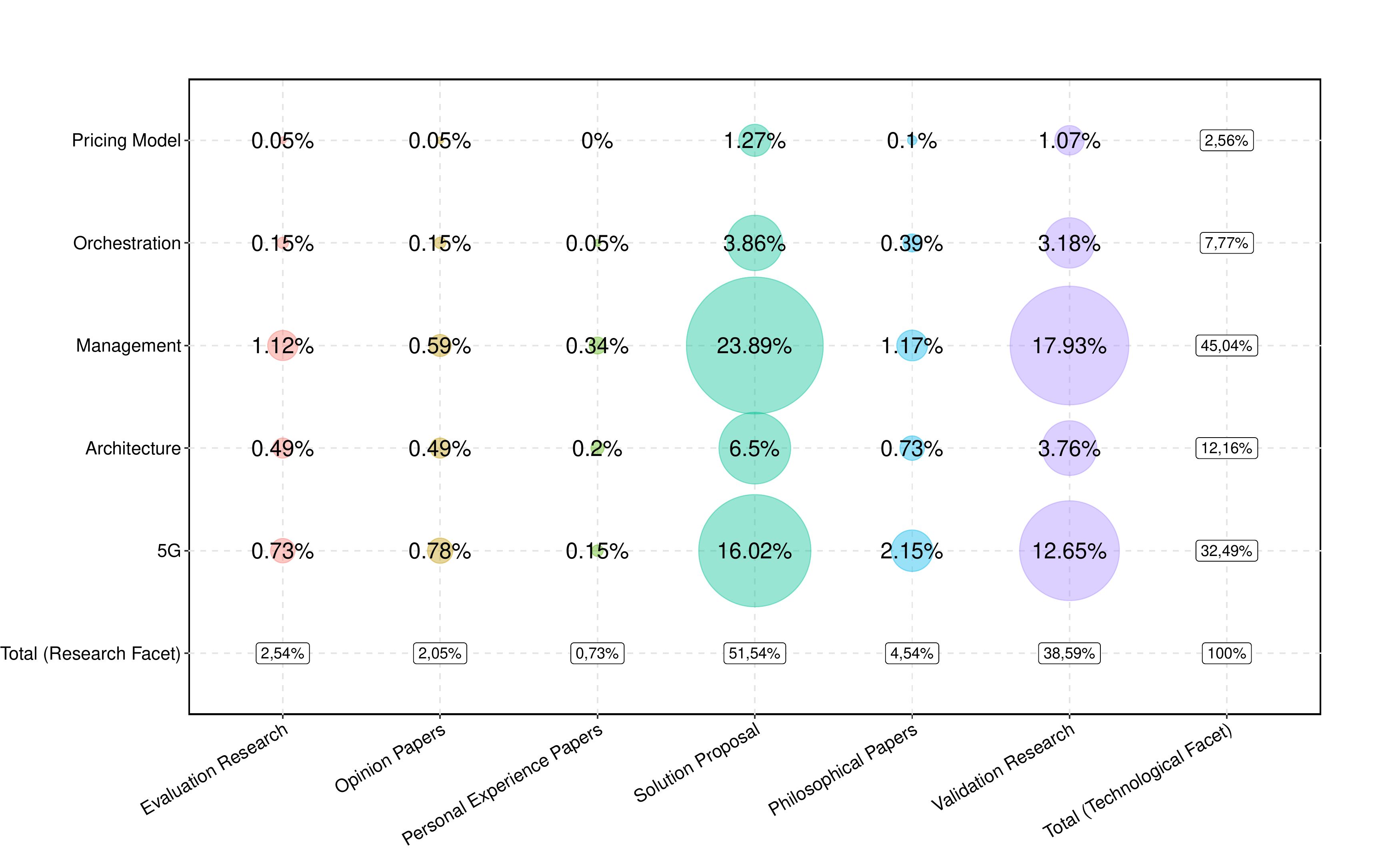}
    \caption{Relationship between research and technological facets (a holistic view).}
    \label{fig:Bubble}
 \end{figure*}

Figure \ref{fig:Bubble} presents one of our main contributions. It highlights the level of relationship between the research facets (RF) defined in \citet{Wieringa:2005} and technological facets (TF) presented in Subsection \ref{sec:Class_proc}. Moreover, several correlations may also be extracted from the figure. 

First of all, the vertical axis (TF) depicts the main research areas being investigated in the last 10 years. We observe that the management area represents 45\% of the papers published, while 5G is in second place with 32,49\%. On the other hand, Orchestration (7,77\%), Pricing Models (2,54\%) and Architecture (12,16\%) are not yet widely explored.

Secondly, let us analyze the results in the horizontal axis (RF). We can notice that 51,54\% of the papers belong to the facet Solution Proposal and 38,59\% to the Validation Research one. As by the definition presented in Subsection \ref{sec:Class_proc}, these two facets are related to small experiments not yet implemented, mostly done in lab. Evaluation Research, which represents solutions implemented in practice, has only 2,54\% of the papers. Easy to conclude that the majority of work (90,13\%) related to CNS is still confined in universities or small setups inside the industry, and should be put into practice in the next upcoming years. 
 
Opinion Papers (2,05\%), Personal Experience Papers (0,73\%) and Philosophical Papers (4,54\%) have still incipient numbers compared to the other facets. Similar behavior can be observed from papers related to Pricing Models. There is only 1.27\% of the indexed papers related to proposed solutions for pricing models in the context of CNS. We suppose that this data demonstrates how difficult it is proposing distinct charging models in the context of CNS, perhaps due to the complexity and distributed nature of the technology.

The next step to observe in Figure \ref{fig:Bubble} is the correlation between TF (vertical axis) and RF (horizontal axis), represented here by bubbles. The biggest bubble is the one between the management area and the solution proposal facet, with 23,89\% of the published papers. The second biggest bubble is the one between management and validation research with 17,93\% of the papers. Lastly, there are still two bubbles that claim our attention: the one between 5G and solution proposal (16,02\%), and another one between 5G and validation research (12,65\%). From these four bubbles, we conclude that the most developed areas of research (\textbf{RQ6}) are the ones related to management and 5G, both still being analysed in small experiments.

On the other hand, we observe areas not well explored by the researchers so far. There are a lot bubbles with percentage numbers between 0 - 5, which means that less effort has been done in those areas. For example, evaluation research facet focused on 5G has only 0.73\% of the papers included. We conjecture that this data demonstrates a 5G research, in the context of CNS, not mature enough to be carried out into practice.

Orchestration, which is a very important feature for CNS \cite{SOUSA:2019} since it encompasses the capability of having a closed-loop, is also still in its infancy having a total of 7,7\% of published papers. From this number, only 3,86\% is in the solution proposal facet and 0,15\% is for evaluation research (related to real implementations). 

Taking all these results into consideration, we are able to answer \textbf{RQ7} by affirming that it is possible to classify papers according to a taxonomy. In this case, the crossing of research and technological facets was used to define a process for classifying studies in the CNS context.

Nonetheless, we observe from Figure \ref{fig:Bubble} that the research methods most used in the literature (in the CNS context) are: solution proposal and validation research. Together, they sum up to 90,13\% of the included papers, which answers \textbf{RQ8}.

\subsection{Inside the Bubbles (In-depth Analysis)} \label{sec:in-depth}

From now until the end of this section, we highlight the behavior inside some bubbles for each technological facet (TF), as defined in Subsection \ref{sec:Class_proc}. The idea is to identify how sub-areas of research inside the facets appear in the indexed papers.

\begin{figure}[ht]
    \centering
     \includegraphics[scale=0.25]{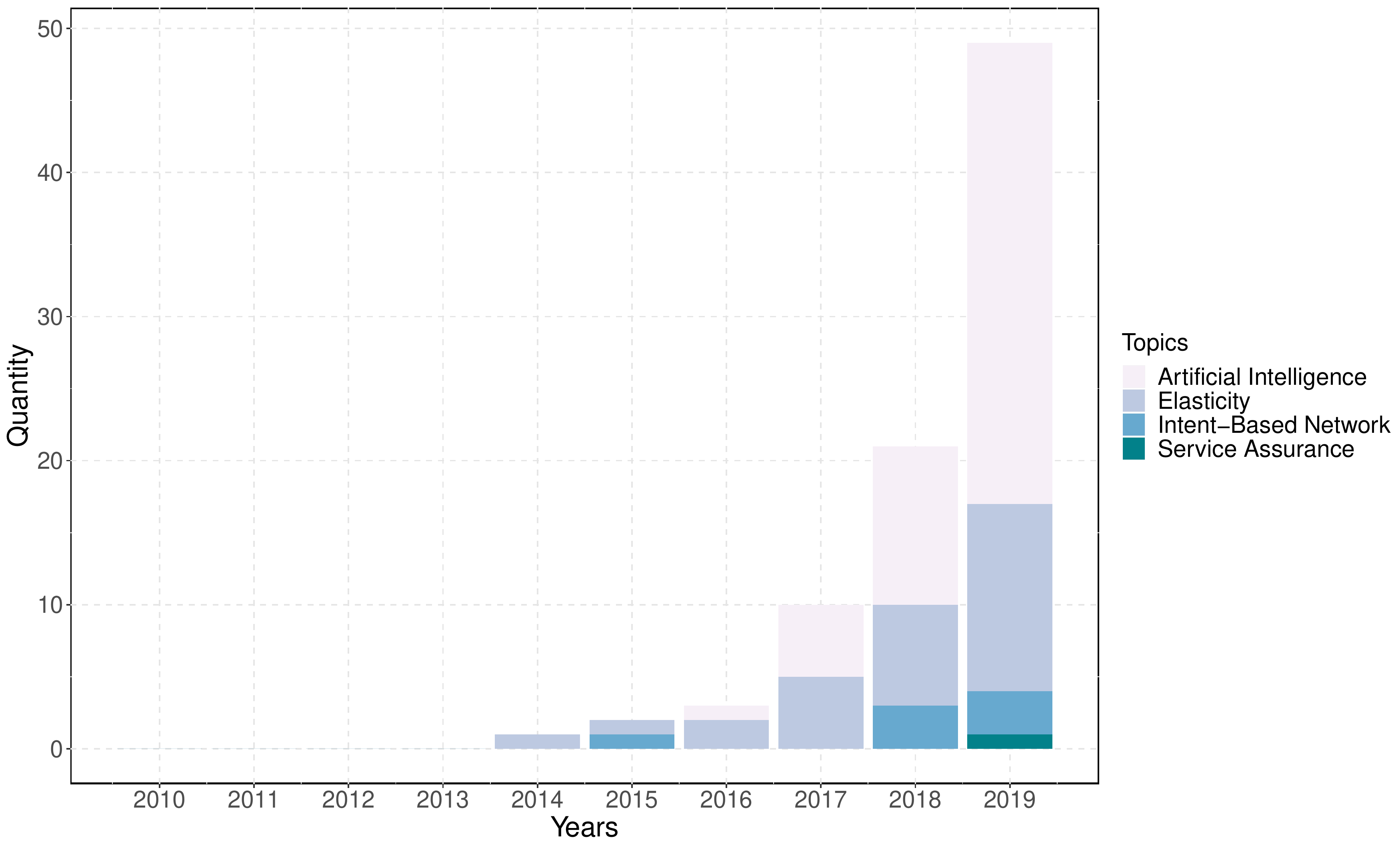}
    \caption{Orchestration.}
     \label{fig:Orch}
\end{figure}

In the literature, Orchestration is treated as a hot topic in CNS context \cite{Gutierrez-Estevez:2018, Gutierrez-Estevez:2019}, but the numbers show that only 7.77\% (Figure \ref{fig:Bubble}) of the papers are focused on this theme. In Figure \ref{fig:Orch}, we detail the results of orchestration and show that a small majority of papers focus on solutions related to the use of Artificial Intelligence (4,19\%). At the same time, Elasticity (2,91\%), Intent-Based Network (0,5\%) and Service Assurance (0,1\%) represent less than a half of the papers included in the Orchestration facet.

Hundreds of recent papers apply machine learning to computer networks. Emerging technologies such as CNS brings a higher level of complexity to the network environment and automation in agile infrastructures (dynamic networks) should open a new range of challenges on applying machine learning in the context of CNS.  In \citet{Chemouil:2019}, AI (Artificial Intelligence) and ML (Machine Learning) are studied in-depth, and the authors concluded that it is necessary to have special care in using these approaches due to the great complexity of data in computer networks.

Elasticity is a feature that can dynamically reduce or add resources to meet tenants need \cite{Yazhou:2016}. In the CNS context, elasticity is a key feature and still an open challenge, due to the need to maintain end-to-end elastic resources across multiple administrative domains \cite{Gutierrez-Estevez:2018}. The CNS architecture must take into account not only the initial requirements, but also the system load, in order to trigger the elasticity process by expanding or reducing the resources available to meet SLA.

Intent-Based Network (IBN) is an approach that captures business intents and translates them into policies that can be automated and applied consistently across the network \cite{IBN:2019}. In this sense, it is related to the provisioning of available resources and establishing new services in the shared infrastructure \cite{Aklamanu:2018}. In the context of orchestration, IBN is a key aspect of services' composition, with distinct requirements that need further study. 

Service Assurance should be the main orchestration component, using dynamic management and monitoring functions in the context of CNS \cite{Xie:2019, Xie2:2019}.

\begin{figure}[ht]
     \centering
     \includegraphics[scale=0.25]{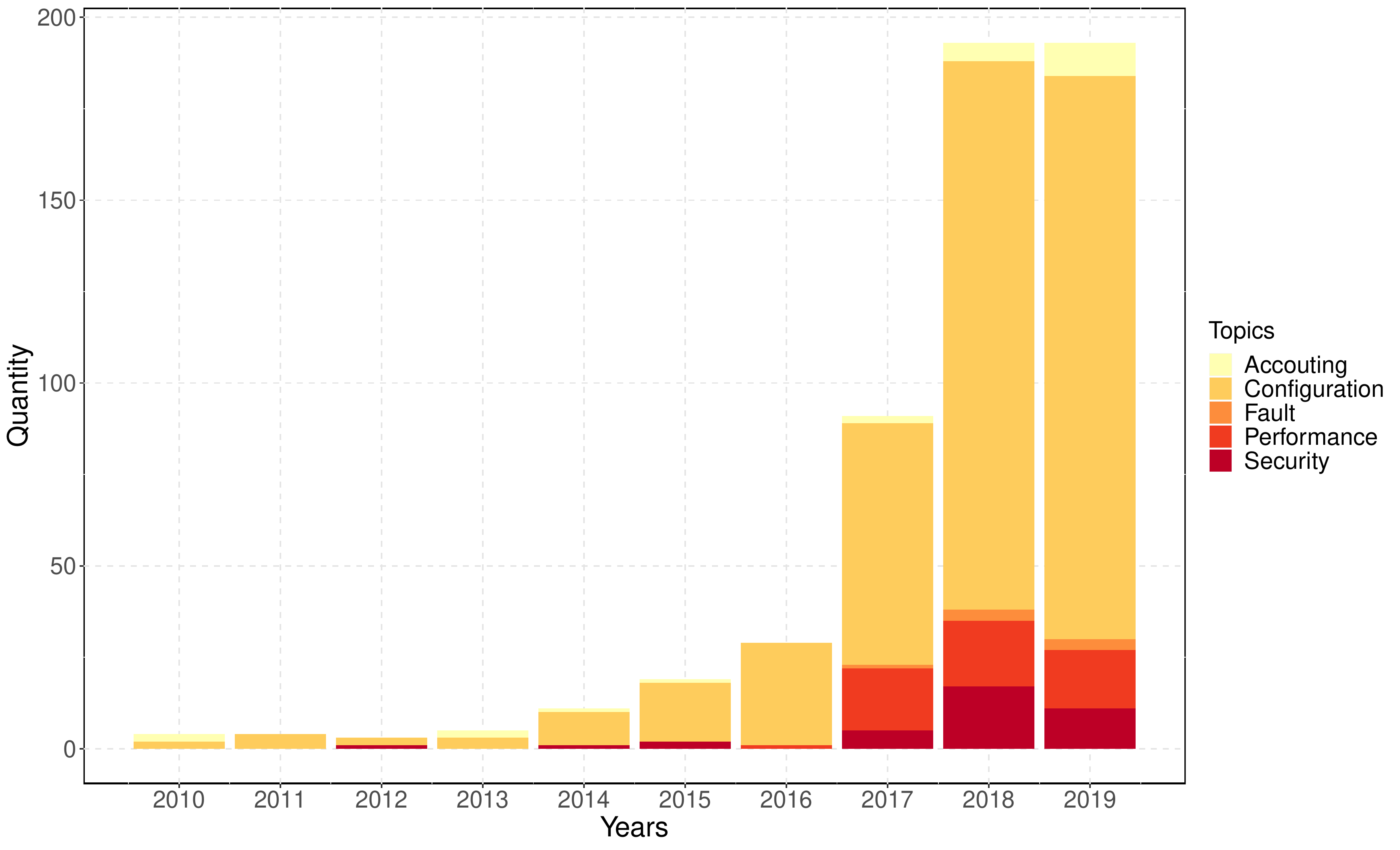}
     \caption{Management.}
     \label{fig:Management}
\end{figure}

In the Management facet, we use the FCAPS \cite{ISO10040:1998} (Fault, Configuration, Accouting, Performance and Security) model to classify papers. Figure \ref{fig:Bubble} shows that between 2009 and 2019 the Configuration subtopic was the main one investigated by the community (35,41\%). Looking at this fact, we believe that enabling technologies such as NFV and SDN have paved the fast development of configuration solutions based on network programmability. Figure \ref{fig:Management} we detail these numbers.

Performance research (4,24\%) in a specific area such as CNS, needs further maturation of operation and configuration proposals. We strongly believe that studies about performance evaluations in the context of CNS should grow in the coming years.

Accounting (1,80\%), Security (3,20\%) and Fault (0,57\%) are currently secondary topics and can be tackled in the near future.

Management and Orchestration are treated by the scientific community as a single topic named MANO. In \citet{Foukas:2017} and \citet{SOUSA:2019}, several unresolved challenges are listed in the CNS area, some of which related to MANO.

According to the 5G specification by 3GPP \cite{3GPP:2019}, network slicing is a key component for enabling multiple services offerings in the same shared infrastructure. A new diversity of network services is expected, from extreme mobile broadband (xMBB) to machine type communications (MTC). The requirements of the services that will be performed on this infrastructure may differ significantly in terms of latency, bandwidth, and many other aspects. In this mapping, the 5G facet obtained the second largest number of indexed papers (32,49\% from Figure \ref{fig:Bubble}). The 5G papers were classified into three areas: RAN, Transport and Core. 

\begin{figure}[ht]
     \centering
    \includegraphics[scale=0.25]{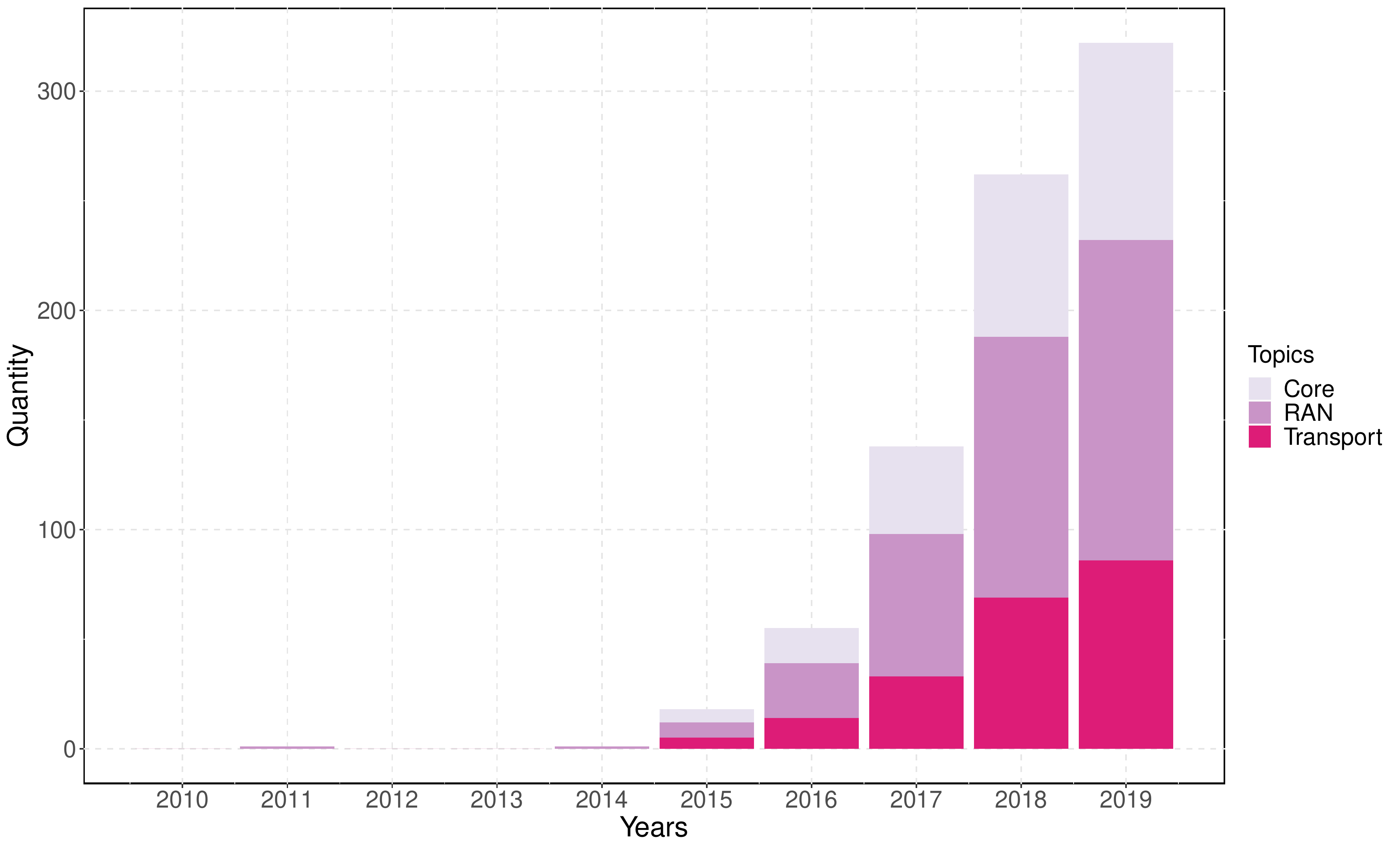}
    \caption{5G.}
    \label{fig:FiveG}
\end{figure}

Figure \ref{fig:FiveG} shows that RAN papers have a quantitative highlight, reaching 14,45\%. We conjecture that RAN in 5G network will encounter a high density of user equipment, which may explain the focus of researchers on this subtopic.

In this mapping, papers related to Transport network reached 9.04\%. On the other hand, papers related to the Core network sum 8.97\%. In the 5G context, Transport network consists of multiple technologies, being a transition point for wireless and optical segments \cite{Giatsios:2017}. In CNS context, the Transport network should aggregate traffic from the edge up to the core and cloud \cite{Oliva:2018}. According to the  \citet{3GPP:2019} specification, 5G core is responsible for connecting the access network through the transport network. 

The complexity of the 5G infrastructure may be a limiting factor for researchers. We suppose that scientific research in transport and core networks requires adequate infrastructure to achieve coherent results. In this case, simulations can be used in order to validate the proposals, however, partnerships with industry can leverage new research possibilities in the 5G context.

\begin{figure}[ht]
     \centering
    \includegraphics[scale=0.25]{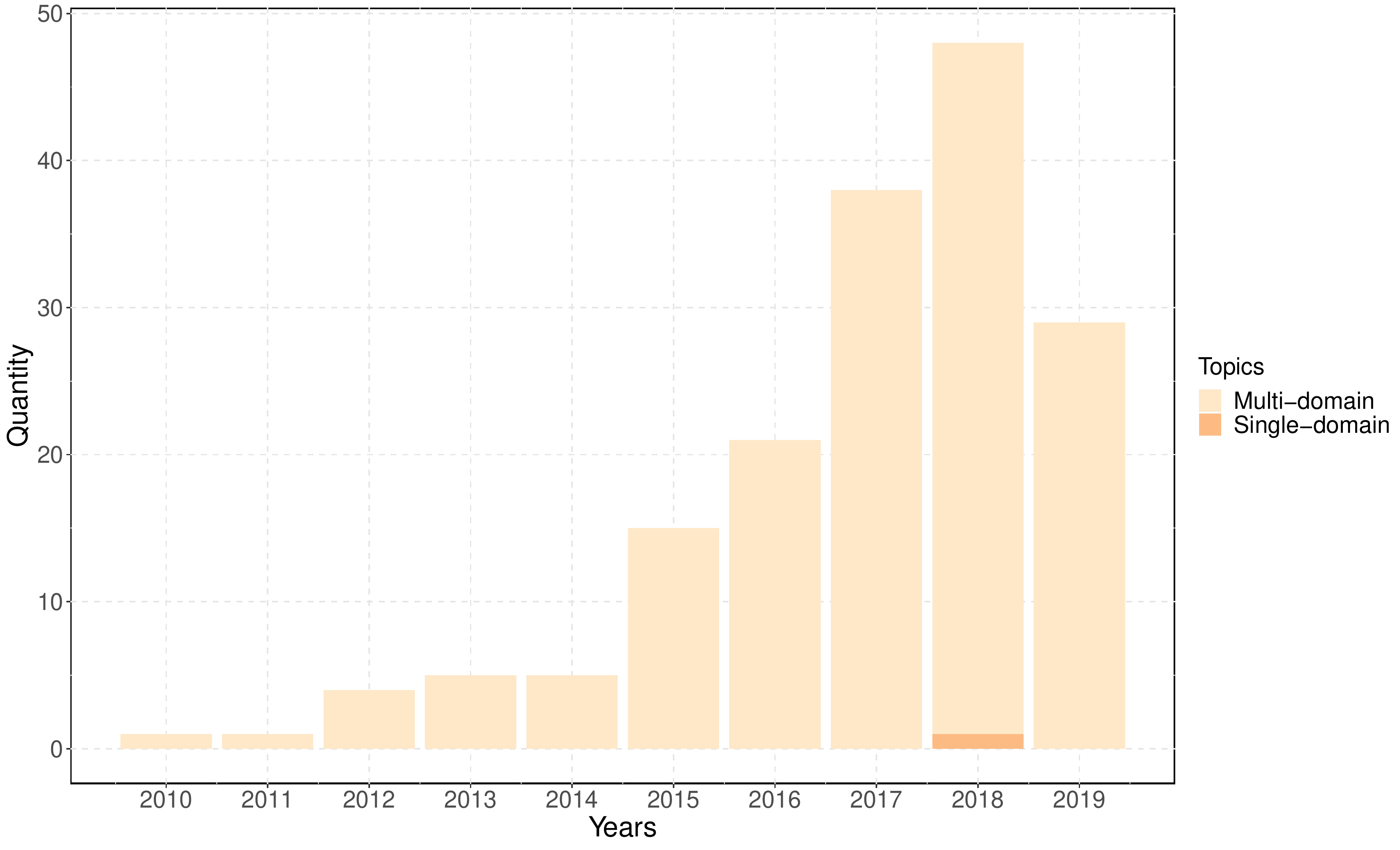}
    \caption{Architecture.}
    \label{fig:Arch}
\end{figure}

Regarding the Architecture facet, Figure \ref{fig:Arch} depicts that most of the studies (almost 100\% from Figure \ref{fig:Bubble}) addressed scenarios with multiple domains. This is an expected result in the context of CNS due to the inherent nature of a Slice, which is to be deployed among different geographically distributed places. As defined by \citet{Galis:2018}, a network slice typically consists of cross-domain components from separate domains in the same or different administrations. These components are applicable to the access network, transport network, core network, and/or edge networks.

\begin{figure}
     \centering
    \includegraphics[scale=0.25]{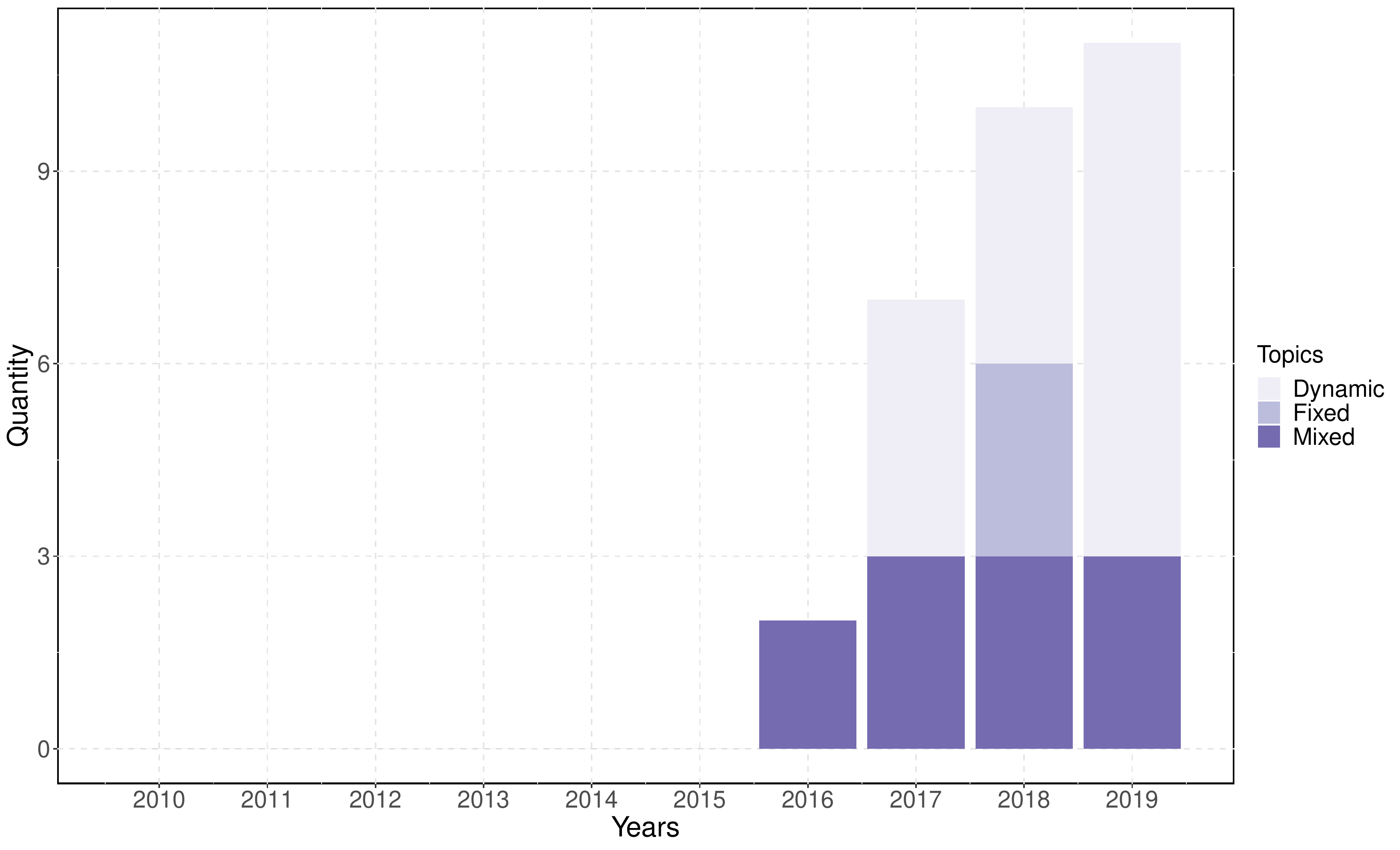}
    \caption{Pricing Models.}
    \label{fig:Billing}
\end{figure}

Creating pricing models on a given technology takes time to mature and to understand the effective demand. In Figure \ref{fig:Bubble}, we observe that Pricing Models is the topic with the least scientific research proposals, reaching 2,54\% of the indexed papers. However, we advocate that defining pricing strategies is a key factor for service adherence. In Figure \ref{fig:Billing}, the Pricing Models were classified in Fixed (0,25\%), Dynamic (1,35\%) or Mixed (0,93\%). 

The dynamics of the environment in the context of CNS, supported by enabling technologies such as SDN/NFV, are reflected in research focusing on dynamic pricing modeling, i.e., it changes according to demand and time. Fixed pricing is probably a simpler way to define pricing models in an environment with dynamic features such as CNS. The mixed pricing model uses a more flexible approach, by combining different types of schemes.

We conjecture that a mixed pricing approach, such as the one used by Amazon\footnote{Available in: https://aws.amazon.com/pt/pricing/.}, is more appropriate in the context of CNS, as the tenant would have more options for choosing. In addition, CNS providers would have more flexibility to negotiate with resource providers in order to form the end-to-end slice.

\section{Open challenges and future directions} \label{sec:OCFD}

Cloud Network Slicing is a hot topic that has considerable depth, due to a complex environment including network, cloud, storage and computing elements. This study showed that, in general, the scientific contributions in the CNS area so far focus on the Management and 5G. On the other hand, issues related to Architecture, Pricing Models and Orchestration models are still incipient.

From the quantitative analysis presented in Section \ref{sec:results} we can list some open challenges and future directions in the context of CNS, thus answering \textbf{RQ9}. Below, we present a short analysis of them.

\subsection{Self-orchestration}

Based on the number of papers published in the technological facet of Orchestration (7,7\%), we observed that this area still needs to be further explored by researchers. In a nutshell, the CNS system must guarantee the execution of the slice making possible adjustments in the availability of resources, thus characterizing the orchestration of the environment. It must take into account not only the initial requirements, but also the system load in order to trigger the elasticity process, scaling up and/or out the available resources to meet the expected service levels. 

In the context of MANO, a recent concept called \textbf{Closed-loop Service Assurance} \cite{Xie:2019} has emerged and can be defined as the capability of having a self-(healing, configuring and optimizing) \cite{IBM:2005} cloud network mechanism, to react to changes in the environment and then triggering actions using autonomic and orchestrated functions. The closed-loop service assurance depends on a sophisticated monitoring system, which collects metrics from the physical and virtual resources, as well as from the services running inside the slices.

Artificial Intelligence algorithms can be used to aid decision making that is part of VNF placement and elasticity of resources \cite{Chemouil:2019}. Some of the candidate machine learning algorithms to be used are the ones related to predictions, such as linear regression.

In fact, we conjecture that the orchestration of management actions in the CNS environment still has open challenges. Proposals based on Artificial Intelligence and Service Assurance point to future directions.

\subsection{Security}

As presented in Subsection \ref{sec:in-depth}, papers focused on Security mechanisms totaled only 3.20\%. It is predicted that a large amount of services will be supported in CNS context, some to be more edge-oriented and others more core cloud-oriented. These will require the complex composition of services and infrastructures, in which the demand for security is also raised \cite{Tao:2018}.

Security is a key topic for the operation of a CNS provider. We believe that setting a comprehensive security policy can drive the composition of security mechanisms along the slice. The challenge is to coordinate the mechanisms that must be supported across multiple domains, containing tenants with distinct operating and security requirements. 

In a virtualized environment, isolation is reached when virtualized and physical components do not have interference at the software level from other components. However, at the end, resources are physical and can be exposed to different components.

Resource isolation is a premise in the context of CNS. \textbf{Multi-level security mechanisms} in heterogeneous environments have not yet been fully developed. Predictive artificial intelligence algorithms can be used to understand the causes of events and behaviors including fault diagnostics and anomaly detection \cite{Chemouil:2019}.

That said, the level of isolation between services that share resources is an open challenge in the context of CNS security. As in other areas, future directions point to the correlation of events using Artificial Intelligence and Blockchain.

\subsection{Pricing models}

With only 2.54\% out of the indexed papers, pricing models in the context of the CNS is the area with the lowest level of development. Creating pricing models is not a simple task if we take into consideration a complex and highly dynamic environment such as CNS. Infrastructure providers may have different pricing models, for example: the network provider uses a fixed pricing model, while the cloud provider uses a mixed pricing model. We have observed that auction models, such as \citet{Habiba:2018}, have been proposed in the 5G context and could be adapted to the CNS scenario.

We believe the CNS provider should continually monitor infrastructure providers and compile the best components from a marketplace \cite{Maciel:2019} in order to offer custom end-to-end slice options for a specific tenant. In this sense, further studies on pricing models are needed to establish the financial viability of a CNS service.

We assume that a future direction is to adopt pricing models already established in cloud providers, such as the ones used by Amazon \cite{Amazon:2019} and Google \cite{Google:2019}.

\subsection{Service deployment}

Although the sub-topic Configuration (a facet of Management) is the one with the largest number of indexed papers (35.41\%), we noted that there are still open challenges in the scope of Service Deployment.

In general terms, a tenant must inform the system about the necessary requirements for the execution of a specific service. The system must be able to interpret the request and reserve resources for a slice by observing the defined SLA \cite{Sciancalepore:2017}. Usually, the service description can be submitted in a high-level language. In this case, the system must be able to translate the high-level description into a set of settings for running the slice (slice resources).

We believe that these requirements can be expressed through the usage of \textbf{Intent-Based Networking} \cite{IBN:2019}. In this context, one of the main challenges here is to map from an abstract high-level service description to slice infrastructure requirements. The mapping process must be performed in several steps, such as service identification, the definition of the initial workload parameters and identification of the restrictions for the execution of the slice. 

That said, mapping resources distributed across multiple domains is an open challenge in the context of CNS. 

\section{Summarization: Research Questions and Answers} \label{sec:RQaA}

We based our study on the definition of nine research questions presented in Subsection \ref{sec:RQ}. These research questions served as motivation for conducting the mapping study in the context of CNS.

In both Section \ref{sec:results} and Section \ref{sec:OCFD}, we answer all of the research questions listed before. In this sense, we present a summary of the research questions and point out the respective answers in Table \ref{tab:RQs}.

\begin{table}[ht] \footnotesize
\renewcommand{\arraystretch}{1.5}
    \centering
    \begin{tabular}{p{3.5cm} c c}
        \hline
        Research Question & Answers & Place\\
        \hline
         1: What are the main companies that make research on CNS? & Table \ref{tab:Industry} & Subsection 4.1  \\
         2: What are the most cited papers in CNS? & Table \ref{tab:Most_cited_papers} & Subsection 4.2 \\
         3: Who are the most cited researchers in CNS? & Table \ref{tab:Researches} & Subsection 4.3  \\
         4: How many publications about CNS have been published in the last 10 years? & Figure \ref{fig:Pub_Years_RA} & Subsection 4.4  \\
         5: What are the top places used so far for publishing papers on CNS? & Table \ref{tab:Places} & Subsection 4.5  \\
         6: What are the most developed areas in CNS? & Figure \ref{fig:Bubble} & Subsection 4.6  \\
         7: Is it possible to classify papers according to a taxonomy? If so, what would it be? & Figure \ref{fig:Bubble} & Subsection 4.6 \\
         8:What are the most frequently applied research methods, and in what study context? & Figure \ref{fig:Bubble} & Subsection 4.6 \\
         9: What are the open challenges in CNS? & Section \ref{sec:OCFD} & Subsections 5.\{1,2,3,4\}  \\
        \hline
        \\
    \end{tabular}
\caption{Summary of RQs and answers.}
\label{tab:RQs}
\end{table}

\section{Conclusions} \label{sec:conclusions}

Exploring a research area can be challenging for young researchers. Correctly discovering the problems that still need to be resolved can be a key factor in the success of a scientific research. In this sense, a systematic mapping study helps in formatting of a research area, allowing the researcher to have a holistic view of it. 

This study made a thorough quantitative analysis of the scientific efforts in the context of Cloud Network Slicing\footnote{All the data used in this work is listed in: http://bit.ly/3b9nlcMPS.}. Evidence from 640 scientific publications were collected in order to understand possible future directions.

In summary, the main results presented in this work are: (1) industry involvement in scientific research was presented quantitatively; (2) the most cited papers are detailed; (3) the most active researchers were listed; (4) the behavior of publications over time has been analyzed and we note that there are still studies to be conducted in the near future; (5) the main publication venues used so far to publish scientific papers in the context of CNS were presented; and (6) a deep and holistic view of the CNS area was highlighted.

Open challenges in CNS area were discussed and future directions were pointed. We conjecture that intent-based networking, service assurance, closed-loop, machine learning and marketplace mechanisms are hot-topics to be investigated in the upcoming years.

\newpage

\begin{landscape}
\clearpage
\onecolumn
\appendix
\begin{adjustwidth}{-3cm}{-3cm}
\section{Structural Analysis} \label{ap:StrucAna}
\begin{scriptsize}
\begin{longtable}{p{2.5cm} p{2.5cm} c p{2.5cm} p{2.5cm} c p{2.5cm}}
        \\
        \hline
        Paper & Problem addressed & Basic approach & Scope & Limitations & Validated & Result\\
        \hline
        \citet{Sunilkumar:2014} & Organize the state-of-the-art in resource management in IaaS clouds. & Survey & Resource management in IaaS environment. & Don't focus on the elasticity approach to a cloud environment distributed by multiple providers. & Yes & Analysis of resource management schemes in IaaS.\\
        \citet{Kokku:2012} &  Design a solution for virtualization of the wireless resources in base stations. & Model & Wireless resources in base stations. & A flow of a client can steal bandwidth allocated to another flow of the same client. & Yes & NVS can virtualize wireless resources in WiMAX networks. \\
        \citet{Akyildiz:2016} & Organize the state-of-the-art in 5G networks. & Survey & 5G networks. & It does not cover solutions that include automation through artificial intelligence to orchestrate the 5G network. & Yes & Analysis of the 10 key enabling technologies in 5G. \\ 
        \citet{Samdanis:2016} & Organize the state-of-the-art of the 3GPP standardization in 5G networks. & Survey & 5G network slice broker. & The Slice Broker does not address competitive conditions for resources. & Yes & Overview of the 3GPP Rel.14 standardization efforts related to multi-service sup-port and network virtualization. \\
        \citet{Richart:2016} & Organize the state-of-the-art in resource allocation and isolation. & Survey & Isolation in virtual wireless networks. & There is no evidence that all studies of resource slicing in virtual wireless networks have been analyzed. & Yes &  Comparative analysis of the existing proposals for wireless resource allocation and isolation. \\ 
        \citet{Foukas:2017} & Organize the state-of-the-art in 5G network slicing. & Survey & 5G network slicing. & It does not address the transport network (edge cloud) in the analysis of the infrastructure layer. & Yes & Evaluation on the maturity of proposals and identification of open research questions. \\
        \citet{Ordonez-Lucena:2017} & Organize the state-of-the-art in network slicing for 5G with SDN/NFV. & Analisys & Network Slicing for 5G with SDN/NFV. & The new directions of research are not discussed in depth. & Yes & Presents an example scenario that combines SDN and NFV technologies to address the realization of network slices.  \\
        \citet{Liang:2015} & Make an integration of the wireless network virtualization and information-centric networking. & Architecture & Wireless network virtualization and information-centric networking & In the proposal, control admission is not supported. & Yes & The performance of backhaul alleviation can be improved. \\
        \citet{Rost:2016} & Discuss the evolution toward a software-defined mobile network control, management, and orchestration. & Analisys & 5G in 3GPP EPS model & The analysis does not show or point to future solutions to the challenges presented. & Yes & Presents technology components and list standards organizations. \\
        \citet{Zhang:2017} & Design of a logical architecture for network-slice-based 5G systems. & Architecture & Mobility and resource management in 5G networks & The architecture does not integrate network slicing with C-RAN, SDN, and NFV. & Yes &  Presents the mechanisms for resource allocation in network-slicing-based 5G networks.\\
        \hline
        \\
    \caption{Structural analysis.}
    \label{tab:Struc_analysis}
\end{longtable}
\end{scriptsize}
\end{adjustwidth}
\clearpage
\twocolumn
\end{landscape}





\end{document}